\documentclass[superscriptaddress,showpacs,preprintnumbers,amsmath,amssymb]{revtex4}
\usepackage{amsfonts}
\usepackage{graphicx}% Include Figure files
\usepackage{dcolumn}% Align table columns on decimal point
\usepackage{bm}% bold math
\usepackage{latexsym}
\usepackage{amsmath}
\begin{document}
\title{Stochastic theory of protein synthesis and polysome:\\ ribosome profile on a single mRNA transcript} 
%\title{Translation by polysome:\\ theory of ribosome profile on a single mRNA transcript} 
\author{Ajeet K. Sharma}
\author{Debashish Chowdhury{\footnote{Corresponding author(E-mail: debch@iitk.ac.in)}}}
\affiliation{Department of Physics, Indian Institute of Technology,
Kanpur 208016, India.}
%\date{\today}%
%%%%%%%%%%%%%%%%%%%%%%%%%%
\begin{abstract}
The process of polymerizing a protein by a ribosome, using a messenger RNA
(mRNA) as the corresponding template, is called {\it translation}. Ribosome
may be regarded as a molecular motor for which the mRNA template serves
also as the track. Often several ribosomes may translate the same (mRNA)
simultaneously. The ribosomes bound simultaneously to a single mRNA
transcript are the members of a polyribosome (or, simply, {\it polysome}).
Experimentally measured {\it polysome profile} gives the distribution of 
polysome {\it sizes}. Recently a breakthrough in determining the instantaneous 
{\it positions} of the ribosomes on a given mRNA track has been achieved and 
the technique is called {\it ribosome profiling} \cite{ingolia10,guo10}.
Motivated by the success of these techniques, we have studied the 
spatio-temporal organization of ribosomes by extending a theoretical model 
that we have reported elsewhere \cite{sharma11}.
This extended version of our model incorporates not only (i) mechano-chemical
cycle of individual ribomes, and (ii) their steric interactions, but also
(iii) the effects of (a) kinetic proofreading, (b) translational infidelity,
(c) ribosome recycling, and (d) sequence inhomogeneities. The theoretical 
framework developed here will serve in guiding further experiments and in 
analyzing the data to gain deep insight into various kinetic processes 
involved in translation.\\

\noindent {\bf Key words}: ribosome traffic, master equation, extremum current 
hypothesis, distance-headway, TASEP.
\end{abstract}
%%%%%%%%%%%%%%%%%%%%%%%%%%
\pacs{87.16.Ac  89.20.-a} 
\maketitle
%%%%%%%%%%%%%%%%%%%%%%%%%%%%%%%%%%%%%%%%%%%%%%%%%%%%%%%%%%%%%%%
\section{Introduction}
%%%%%%%%%%%%%%%%%%%%%%%%%%%%%%%%%%%%%%%%%%%%%%%%%%%%%%%%%%%%%%%

Ribosome \cite{spirin00,spirin02,frank06} is a macromolecular complex and 
operates as one of the essential intracellular machines \cite{frank11} 
that participate in gene expression in all living cells \cite{alberts,lodish}.   
More specifically, it polymerizes a protein that is a linear hetero-polymer 
consisting of amino-acid monomers each of which is linked to the next one by 
a peptide bond. Therefore, growing protein is also called a polypeptide. For 
the synthesis of a protein, a messenger RNA (mRNA) serves as the template; 
the sequence of the amino acid species in the protein is determined by that 
of the codons (triplets of nucleotides) on the corresponding mRNA template. 
This process is called {\it translation}. Translation by every ribosome goes 
through three main stages: (i) {\it initiation}, (ii) {\it elongation}, and 
(iii) {\it termination}. The {\it start} and {\it stop} codons mark the 
positions on the template mRNA where initiation and termination of 
translation take place.

During the elongation stage, at every codon, the amino acid monomer required 
for elongating the protein is supplied by an incoming tRNA molecule; the 
correct amino acid monomer is carried by those tRNA whose anti-codon is 
complementary to the codon. The machinery of translation deploys a quality 
control mechanism which screens the incoming tRNA through a multi-step  
selection process. However, in spite of this stringent selection process, 
occasionally an incorrect amino acid may escape rejection by the quality 
control system; a {\it translational error} results if the growing protein 
incorporates an incorrect amino acid monomer thereby lowering the {\it 
fidelity} of translation. In any case, after the termination, a ribosome 
is partly disassembled. These parts can reach near the start codon by 
diffusion in the surrounding aqueous medium. A ribosome can be assembled 
more quickly from these parts than from basic constituents. Moreover, in 
case the start and the stop codons are close to each other because of the 
loop formation by the mRNA, diffusive transfer of the parts of the ribosome 
from the stop codon to the start codon can be quite rapid leading to a 
faster {\it recycling} of the ribosomes \cite{hirokawa06}. Rarely elongation 
process is aborted because of the premature detachment of the ribosome from 
the mRNA track. Furthermore, often several ribosomes translate the same mRNA 
transcript simultaneously, each polymerizing a distinct copy of the same 
protein. Because of the superficial similarities with vehicular traffic 
on a given stretch of a highway \cite{css00,scn10,polrev05}, the simultaneous 
collective translation of a mRNA by several ribosomes is sometimes referred 
to as ribosome traffic. The ribosomes bound simultaneously to a single mRNA 
transcript are the members of a polyribosome (or, simply, {\it polysome}) 
\cite{warner62,warner63,rich04,noll08}. 
Because of the mutual hindrance of the ribosomes, the overall rate of 
protein synthesis is expected to attain a maximum at an optimum mean separation 
between the ribosomes. Finally, the ongoing production and decay of mRNA 
transcripts and various feedback loops in gene expression also control the 
rate of protein synthesis.

It would be desirable to capture all the processes mentioned above within a 
single theoretical model of translation. However, it is extremely unlikely 
that such a model can be analyzed analytically. Therefore, the aim of this 
paper is more modest. Here we extend our earlier model \cite{sharma10,sharma11}.
capturing (i) the mechano-chemical cycle of individual ribomes, and (ii) 
their steric interactions, as well as (iii) the effects of 
(a) quality-control mechanisms, (b) translational error, (c) ribosome 
recycling, and (d) sequence inhomogeity of the mRNA. 

The overall rate of synthesis of proteins is a key quantity in any model 
of translation. However, the main focus of our theoretical study here are 
the size of the polysome and the spatial distribution of ribosomes on a 
mRNA. We identify the different parameter regimes of our theoretical model 
and characterize these in terms of the average density of the polysome and 
the overall average rate of synthesis of proteins from a single mRNA 
transcript. Moreover, going beyond the scope of all the previous 
theoretical works on this topic, we predict the nature of the {\it 
fluctuations} in the spatial organizations of the ribosomes which throws 
light on the fluctuations in the size of ribosome clusters on a given 
mRNA transcript.  

In this paper we also suggest a new experiment for testing our theoretical 
predictions on the statistical properties of polysomes. 
Traditional technique of polysome profiling \cite{arava03,mikamo05} provide 
the number of ribosomes bound to a mRNA, but not their individual position 
at the instant when translation was stopped by the experimental protocol. 
An improved version of this technique, called {\it ribosome density mapping} 
\cite{arava05}, provides more detailed information on the numbers of 
ribosomes associated with specified {\it segments} of a particular mRNA 
by carrying out site-specific cleavage of the mRNA transcript. The 
results obtained using these techniques are often adequate for getting a 
qualitative indicator of the translational activity. However, the ribosomes 
are not expected to be uniformly distributed on a mRNA because of the 
stochasticities in the steps of the mechano-chemical cycles of these cyclic 
machines. These stochasticities arise from (i) {\it intrinsic} fluctuations 
in biochemical processes at low copy numbers of the molecules, and (ii) 
{\it extrinsic}  fluctuations arising from the sequence inhomogeneity of 
the mRNA. The most detailed picture of the translational activity has been 
obtained by a recently developed technique, called {\it ribosome profiling} 
\cite{ingolia10,guo10}. 
For testing some of our theoretical predictions, the older technique of 
polysome profiling is adequate whereas for testing the other new results 
ribosome profiling would be necessary.

This paper is organized as follows: we introduce our model and write 
down the master equations for the stochastic kinetics of this model in 
section II. In section III we solve the master equations in the steady 
state under periodic boundary conditions to calculate the overall rate 
of protein synthesis. The results demonstrate the effects of steric 
hindrance caused by congestion ribosome traffic. The spatio-temporal 
organization of the ribosomes in different parameter regimes correspond 
to the different non-equilibrium phases on the ``phase diagrams'' which 
we plot in section IV. The instantaneous spatial distribution of the 
ribosomes on a single mRNA is also characterized in terms of some 
quantitative measures which we introduce in  section V where we also 
explore the effects of sequence inhomogeneities of mRNA. Finally, in 
section VI, the main results are summarized and important conclusions 
are drawn.

%%%%%%%%%%%%%%%%%%%%%%%%%%%%%%%%%%%%%%%%%%%%%%%%%%%%%%%%%%%%%%%
\section{Model}
%%%%%%%%%%%%%%%%%%%%%%%%%%%%%%%%%%%%%%%%%%%%%%%%%%%%%%%%%%%%%%%

The kinetic models of translation can be divided into three categories. 
Translation is just a single step in the broader context of gene 
expression. However, in most of the kinetic models of gene expression 
\cite{zhdanov11}, the details of the mechano-chemistry of individual 
ribosomes as well as their mutual steric interactions are ignored.  
The rates of synthesis and degradation of proteins are captured 
usually in these models by two rate constants without any mechanistic 
details of these two processes. We are not concerned with a global 
picture of gene expression in this paper and, therefore, such kinetic 
models will not be discussed further here. 

There are models of translation which are intended to describe various 
key aspects of the stochastic mechano-chemical kinetics of only a 
single ribosome. In contrast, another class of models of translation 
is motivated by the polysome formation. Most of these models capture 
the effects of entire mechano-chemical cycle by a single parameter. 
These models focus mainly on the effects of mutual steric interactions 
of the ribosomes on the overall rate of protein synthesis. In this 
section we develop a model by capturing both these aspects of 
translation, namely, details of single-ribosome mechano-chemistry 
and the effects of steric interactions among the ribosomes on the 
same mRNA transcript. However, for the  convenience of comparison of 
our work with earlier works, we summarize the main features of the 
TASEP-type models in the next subsection.

%%%%%%%%%%%%%%%%%%%%%%%%%%%%%%%%%%%%%%%%%%%%%%%%%%%%%%%%%%%%%%%
\subsection{TASEP-type models}
%%%%%%%%%%%%%%%%%%%%%%%%%%%%%%%%%%%%%%%%%%%%%%%%%%%%%%%%%%%%%%%

Totally asymmetric simple exclusion process (TASEP) 
\cite{derrida,schuetz}
is one of the simplest models of interacting self-propelled particles; 
it is used extensively for understanding the generic features of 
non-equilibrium steady-states of interacting systems. TASEP and its 
various extensions exhibit interesting dynamical phase transitions 
\cite{mukamel00}.  
For many years, various biologically motivated extensions of TASEP 
\cite{macdonald68,macdonald69,lakatos03,shaw03,shaw04a,shaw04b,chou03,chou04,zia11,zouridis08,romero10}
have been used to model ribosome traffic. In the TASEP-based models 
of ribosome traffic (see ref.\cite{zia11} for a recent review) 
each lattice site represents a single codon. Since a ribosome is much 
larger than a single codon, each ribosome is represented by a hard 
rod that covers ${\ell}$ (${\ell} > 1$) sites simultaneously. But, 
the allowed step size of a rod is one lattice site (i.e., one codon). 
This extended version of TASEP for hard rods will be referred to as 
${\ell}$-TASEP. As long as a site remains covered by a ribosome, it 
is inaccessible to the other rods. The entry of a rod from one end 
(at a rate $\alpha$) and its eventual exit from the other end (at a 
rate $\beta$) model the {\it initiation} and {\it termination} stages 
of translation by a ribosome. 

The steps of the mechano-chemical cycle of individual ribosomes during 
the {\it elongation} stage were not captured explicitly in the simple 
TASEP-type models; instead, one single ``hopping'' parameter was used 
to describes the rate of translation of one codon. Moreover, these 
TASEP-type models neither incorporate any mechanism for selecting 
specifically the correct amino acid monomer, nor do these allow for the 
possibility of translational error. Therefore, such TASEP-type models 
are too simple to account for the effects of various mechano-chemical 
processes on the statistical properties of polysomes.

%%%%%%%%%%%%%%%%%%%%%%%%%%%%%%%%%%%%%%%%%%%%%%%%%%%%%%%%%%%%%%%
\subsection{Our model: unification of single-ribosome mechano-chemistry and TASEP}
%%%%%%%%%%%%%%%%%%%%%%%%%%%%%%%%%%%%%%%%%%%%%%%%%%%%%%%%%%%%%%%

In recent years, progressively more realistic models of translation have 
been developed \cite{ciandrini,basu07,gccr09,sharma10,sharma11} 
and several analytical results have been derived. Using the most recent 
version of this model \cite{sharma11}, some statistical properties of 
single ribosome have been derived analytically \cite{sharma11}. Here we 
extend this model even further to capture some features of translation 
which were not included in its earlier version. Using this extended 
version of our model of translation, we make experimentally testable 
predictions on the dependence of the statistical properties of the 
polysomes on the various mechano-chemical processes involved in 
translation.  

Each ribosome consists of two subunits which are designated as ``large'' 
subunit and ``small'' subunit, respectively. The translation of the 
genetic message encoded in the codon is carried out by the small subunit 
while the elongation of the polypeptide, by the formation of a peptide bond 
between it and the incoming amino acid, takes place in the large subunit. 
The function of the two subunits is coordinated by the tRNA molecules.
There are three binding sites for a tRNA on each ribosome; these sites 
are designated as E,P and A. An incoming tRNA binds with an A site. 
The amino acid carried by a tRNA is linked to the growing polypeptide 
by a peptide bond while the tRNA is bound to the P site. Finally, the 
denuded tRNA exits from the ribosome from the E site. During the 
elongation stage, in each complete mechano-chemical cycle, the ribosome 
steps forward on the template mRNA by one codon while, simultaneously, 
the polypeptide gets elongated by one amino acid. These processes 
are captured explicitly in the our kinetic model.

\begin{figure}[ht]
\begin{center}
\includegraphics[width=0.75\columnwidth]{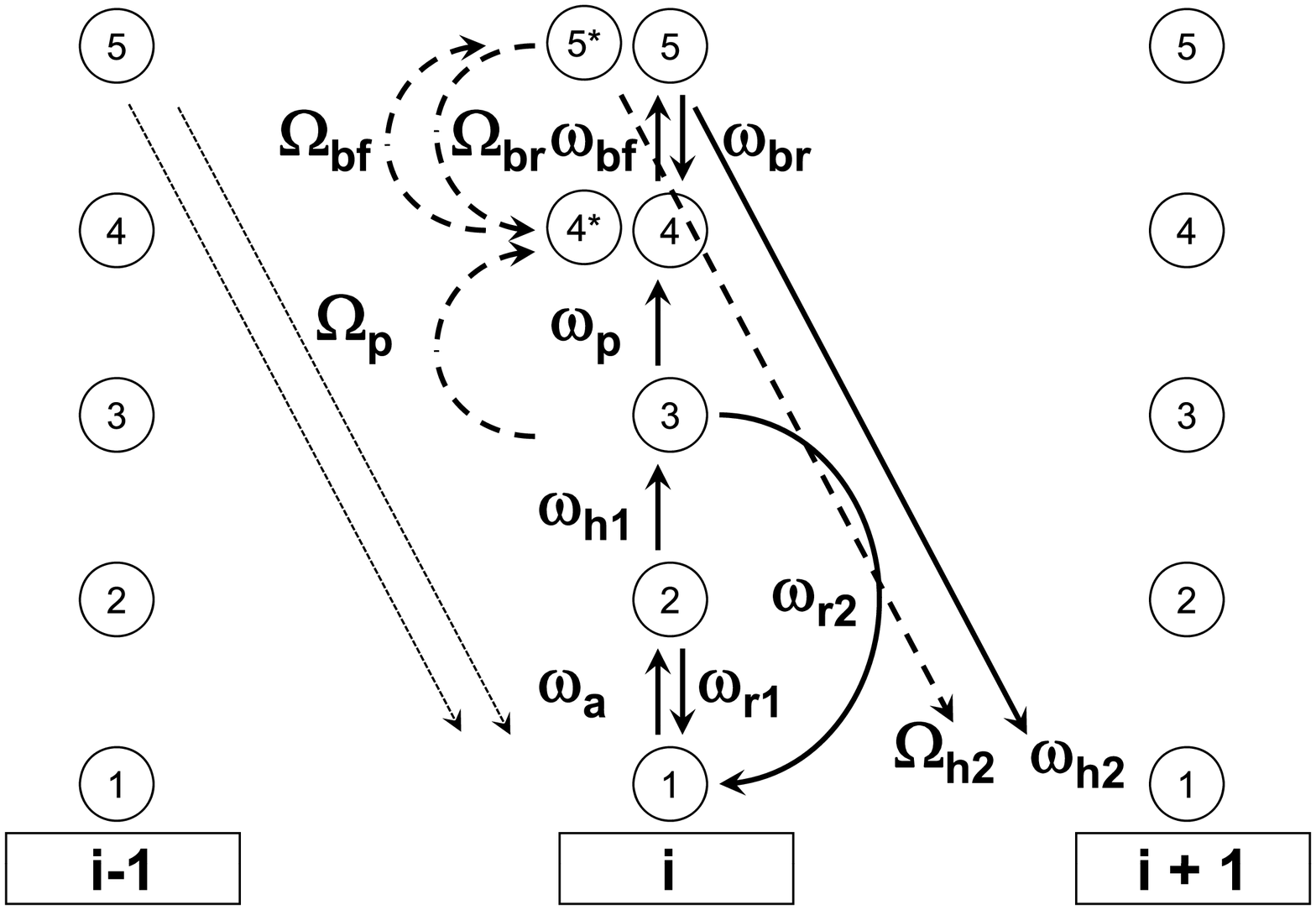}
\end{center}
\caption{} 
\label{fig-fig2}
\end{figure}

The distinct mechano-chemical states in our model and the allowed 
transitions among these states are shown schematically in fig.\ref{fig-fig2}.
At the beginning of each cycle, the system is in state 1 where the sites 
E and A are empty while the site P is occupied by a tRNA that has just 
contributed its amino acid to the growing polypeptide. A tRNA charged 
with an amino acid is called a aminoacyl tRNA (aa-tRNA). At this stage, an 
aminoacyl tRNA, bound to an elongation factor Tu (EF-Tu) and a molecule 
of Guanosine triphosphate (GTP) enters and binds with the A site on the 
Ribosome at the site A. This process takes place with rate $\omega_a$ 
which causes transition of the system to the chemical state 2. Thereafter 
non-cognate tRNAs are rejected, and the system reverts back to state 1, 
 with rate $\omega_{r1}$, through a quality control mechanisms based on 
the free energy of codon-anticodon matching. 

However, the free-energy difference between the cognate and near-cognate 
tRNAs is too small to distinguish between them. Therefore, usually, 
near-cognate tRNAs are not rejected at this stage. A second stage of 
quality control, called {\it kinetic proofreading} \cite{hopfield,ninio}, 
is then activated. GTP, which is bound to the aa-tRNA, is hydrolyzed 
to GDP by EF-Tu and this process is described by the transition from the 
state $2$ to the state $3$. At this stage, barring a few exceptional cycles, 
the 
near cognate tRNAs are rejected from chemical state 3 which drives the 
system back to the chemical state 1; this happens with rate constant 
$\omega_{r2}$. 

Although most often the noncongnate and near cognate tRNAs are rejected 
by the two-stage selection process, still occasionally the quality control 
system fails to reject an incorrect (non-cognate or near-cognate) tRNA. 
Consequently, there is a small, but non-vanishing probability, of a 
translational error when the growing polypeptide elongates by the 
formation of a peptide bond with an incorrect amino acid. In our model, 
the incorporation of incorrect amino acid leads to a branched pathway: 
in contrast to the transition $3 \to 4$ along the correct pathway, the 
wrong pathway proceeds by the transition $3 \to 4^{*}$. Arrival of 
another elongation factor called EF-G, alongwith a molecule of GTP also 
takes place at this stage. The transition $4 \to 5$ (or, $4^{*} \to 5^{*}$) 
is reversible and essentially a Brownian rotation of the two subunits 
relative to each other. This spontaneous Brownian rotation drives the 
two tRNA molecules back-and-forth between the classical P/P, A/A state 
and the hybrid E/P, P/A state \cite{agir09}. The rate constants for the forward 
and backward Brownian rotations are denoted by $\omega_{bf}$ and 
$\omega_{br}$, respectively, along the correct pathway whereas the same 
transitions along the wrong pathway take place with the rates 
$\Omega_{bf}$ and $\Omega_{br}$, respectively. Finally, hydrolysis of 
GTP by EF-G drives the process of {\it translocation} at the end of which 
the two tRNA molecules are positioned at the E and P sites while the 
ribosome finds itself poised to translate the next codon; the denuded 
tRNA molecule makes an exit from the E site. The transition $5 \to 1$ 
and $5^{*} \to 1$ take place with the rates $\omega_{h2}$ and $\Omega_{h2}$, 
respectively. The completion of the full cycle elongates the protein by 
one amino acid (by correct amino acid along one pathway and by an incorrect 
amino acid along another branch) and translocates the ribosome by one codon 
on the template mRNA (For further details, see ref.\cite{agir09}).

Since our model allows the possibility of translational error, we 
define \cite{sharma10} the {\it fidelity} $\phi$ of translation by the 
fraction of the incorporated amino acids which are correct, i.e., 
\begin{equation}
\phi=\omega_p/(\omega_p+\Omega_p)
\end{equation}
In our model, the mRNA track is represented by a one-dimensional lattice 
where each of the total $L$ sites corresponds to a single codon. 
The length of a ribosome is denoted by ${\ell}$ in the units of the 
length of a single codon. A ribosome can move forward by only one 
site (i.e., one codon) at a time. We use the convention that the 
{\it leftmost} site covered by a ribosome is the one that is being 
translated by it; the leftmost site covered by a ribosome is also 
used in our formulation to denote the position of a ribosome. Thus, 
throughout this paper, we follow the convention that, at any instant 
of time, a ribosome ``covers'' ${\ell}$ sites but ``occupies'' only 
the leftmost of these ${\ell}$ sites. 

According to our notation, the status of coverage of a site is 
denoted by $0$ and $1$; $0$ represents an unoccupied lattice site 
whereas $1$ represents a covered site. Since many ribosomes move 
simultaneously on the same track they also interact with each other. 
The simplest form of interaction would be mutual exclusion: if the 
$i$th site is ``occupied'' by one ribosome, then all the ${\ell}$ 
sites from $i$ to $i+{\ell}-1$ are ``covered'' by it and, therefore, 
none of the these ${\ell}$ sites are accessible to any other ribosome 
at that instant of time. Moreover, a ribosome occupying the position 
$i$ can move forward if, and only if, the site $i+{\ell}$ is not 
simultaneously occupied by another ribosome. In our notation, 
the symbol $P(\underbrace{1..............1}_{\ell}|\underline{0})$ 
represents the conditional probability that, given an uncovered site, 
there will be successive $\ell$ adjacent sites to its left all of 
which are covered simultaneously by a single ribosome. Similarly,  
$P(\overbrace{\underline{1..............1}}^{\ell}|~0)$ is the 
conditional probability of finding a empty site $j$, given that the 
successive $\ell$ sites on its left are covered by a ribosome. 

Using the same notation, we now define $Q(\underline{i}|i+{\ell})$  
as the conditional probability that the site $i+{\ell}$ is {\it not} 
occupied by another ribosome, given that the site $i$ is occupied by 
a ribosome, Similarly, given that the site $i$ is occupied by a 
ribosome, the probability that the site $i-{\ell}$ is not occupied 
by another ribosome is given by the conditional probability 
$Q(i-1|\underline{i})$. It is straightforward to show that 
\cite{basu07,gccr09} 
\begin{equation}
Q=(1-\rho\ell)/(1+\rho-\rho\ell)
\end{equation}
Where $\rho=N/L$ is the number density of the Ribosome on the $mRNA$ track.
Following the same prescription that one of us (DC), and his collaborators, 
used in earlier simpler models of ribosome traffic \cite{basu07,gccr09}, 
we multiply the rate constants $\omega_{h2}$ and $\Omega_{h2}$ by Q 
because a ribosome ``feels'' the mutual exclusion only when it tends to 
move forward to the next codon. 

By the symbol $P_{\mu}(i,t)$ we denote the probability of finding the 
ribosome in the $\mu$th chemical state while it occupies the site $i$ 
at time $t$.  The master equations for the probabilities 
$P_{\mu}(i,t)$ are 
\begin{equation}
dP_1(i,t)/dt = -\omega_a P_1(i,t)+\omega_{r1} P_2(i,t)+ \omega_{r2} P_3(i,t)+\omega_{h2}QP_5(i-1,t)+\Omega_{h2}QP_5^*(i-1,t)
\label{eq-m1}
\end{equation}
\begin{equation}
dP_2(i,t)/dt=\omega_a P_1(i,t)-(\omega_{r1}+\omega_{h1}) P_2(i,t)
\label{eq-m2}
\end{equation}
\begin{equation}
dP_3(i,t)/dt=\omega_{h1} P_2(i,t)- (\omega_p+\Omega_p+\omega_{r2})P_3(i,t)
\label{eq-m3}
\end{equation}
\begin{equation}
dP_4(i,t)/dt=\omega_p P_3(i,t)-\omega_{bf} P_4(i,t)+\omega_{br} P_5 (i,t)
\label{eq-m4}
\end{equation}
\begin{equation}
dP_5(i,t)/dt=\omega_{bf} P_4(i,t)-(\omega_{h2}Q+\omega_{br}) P_5(i,t)
\label{eq-m5}
\end{equation}
\begin{equation}
dP_4^*(i,t)/dt=\Omega_p P_3(i,t)-\Omega_{bf} P_4^*(i,t)+\Omega_{br} P_5^* (i,t)
\label{eq-m6}
\end{equation}
\begin{equation}
dP_5^*(i,t)/dt=\Omega_{bf} P_4^*(i,t)-(\Omega_{h2}Q+\Omega_{br}) P_5^*(i,t)
\label{eq-m7}
\end{equation}
Equations (\ref{eq-m1})-(\ref{eq-m5}) correspond to the equations 
(52)-(56) of ref.\cite{gccr09}. However, there are some additional 
terms in (\ref{eq-m1})-(\ref{eq-m5}) because of (i) the kinetic 
proofreading, (ii) translational because of wrong amino acid 
selection, and (iii) reversible nature of the transition between 
the states $4$ and $5$. Moreover, the precise interpretations of 
the respective states were slightly different in ref.\cite{gccr09} 
(see ref.\cite{gccr09} for the detailed interpretations of the 
states and transitions between the states). 

Note that the normalization condition for the probabilities is 
\begin{equation}
\sum_{\mu=1}^{5} P_{\mu}(i,t) + P_{4}^{*}(i,t) + P_{5}^{*}(i,t) = \rho
\label{eq-norm}
\end{equation} 

Molecular mechanisms that lead to mistranslation have been under 
intense investigation for decades (see \cite{reynolds10} for a 
recent review). However, to our knowledge, none of the earlier 
models, including that developed in ref.\cite{gccr09}, provides 
a mathematical framework to treat the mechanisms of mistranslation 
analytically. 

%%%%%%%%%%%%%%%%%%%%%%%%%%%%%%%%%%%%%%%%%%%%%%%%%%%%%%%%%%%%%%%
\section{Rate of protein synthesis: effects of hindrance in ribosome traffic}
%%%%%%%%%%%%%%%%%%%%%%%%%%%%%%%%%%%%%%%%%%%%%%%%%%%%%%%%%%%%%%%

We solve the equations (\ref{eq-m1})-(\ref{eq-m7}) in the steady 
state under normalization condition (\ref{eq-norm}) and imposing 
periodic boundary conditions (PBC). 
Using the steady-state solutions for $P_{\mu}$ we get the 
following expression for the steady-state flux
\begin{equation}
 J_{PBC}=(P_5\omega_{h2}+P_5^*\Omega_{h2})Q =\rho K_{eff}\biggl(1+(\Omega_p/\omega_p)\biggl)
\end{equation}
where 
\begin{eqnarray}
K_{eff}^{-1} &=& \omega_{a}^{-1}\biggl(1+(\omega_{r1}/\omega_{h1})\biggr)\biggl(1+(\omega_{r2}/\omega_{p})\biggr)
+\omega_{h1}^{-1}\biggl(1+(\omega_{r2}/\omega_{p})\biggr)+\omega_{p}^{-1}
+\omega_{bf}^{-1}\biggl(1+(\omega_{br}/\omega_{h2} Q)\biggr) + (\omega_{h2} Q)^{-1}\nonumber \\
&+&\biggl(\Omega_p/\omega_p\biggr)\biggr[\omega_{a}^{-1}\biggl(1+(\omega_{r1}/\omega_{h1})
\biggr)+\omega_{h1}^{-1}
+\Omega_{bf}^{-1}\biggl(1+(\Omega_{br}/\Omega_{h2} Q)\biggr)+(\Omega_{h2} Q)^{-1}\biggr]
\label{a1}
\end{eqnarray}
Separating out the $Q$-dependent and $Q$-independent parts, $K_{eff}$ 
can be re-expressed as
\begin{equation}
K_{eff}^{-1}=k_1^{-1}+(k_2 Q)^{-1}
\end{equation}
where
\begin{eqnarray}
k_1^{-1} &=&
\omega_{a}^{-1}\biggl(1+(\omega_{r1}/\omega_{h1})\biggr)\biggl(1+(\omega_{r2}/\omega_{p})\biggr) 
+\omega_{h1}^{-1}\biggl(1+(\omega_{r2}/\omega_{p})\biggr)
+\omega_{p}^{-1}+\omega_{bf}^{-1} \nonumber \\
&+&\biggl(\Omega_p/\omega_p\biggr)\biggl[\omega_a^{-1}\biggl(1+(\omega_{r1}/\omega_{h1})\biggr)+\omega_{h1}^{-1}+\Omega_{bf}^{-1}\biggr]
\end{eqnarray}
and
\begin{equation}
k_2^{-1}=\omega_{h2}^{-1}\biggl(1+(\omega_{br}/\omega_{bf})\biggr)+\Omega_{h2}^{-1}\biggl(1+(\Omega_{br}/\Omega_{bf})\biggr)(\Omega_p/\omega_p)
\label{eq-k2}
\end{equation} 
Note that when simultaneously $\omega_{r2} \to 0$, $\Omega_{p} \to 0$, 
$\omega_{br} \to 0$, the expressions for $k_{1}$ and $k_{2}$ reduce 
to $K_{eff}$ and $\omega_{h2}$, respectively.

So the $J_{PBC}$ is given by  
\begin{equation}
J_{PBC}=\biggl[k_2\rho(1-\rho {\ell})\biggl(1+(\Omega_p/\omega_p)\biggr)\biggr]/\biggl[\biggl(1+(k_2/k_1)\biggr)(1-\rho {\ell})+\rho\biggr].
\label{eq-jpbc}
\end{equation}
Since no premature detachment of ribosomes are allowed in our model, 
the flux of the ribosomes is also the total rate of protein synthesis.  
In the special case 
$\omega_{r2} = \omega_{br} = \Omega_{p} = \Omega_{bf} = \Omega_{br} = \Omega_{h2} = 0$
the expression (\ref{eq-jpbc}) reduces to the corresponding expression 
for flux derived in ref.\cite{gccr09}.
The expression (\ref{eq-jpbc}) for $J_{PBC}(\rho)$ exhibits a single 
{\it maximum} which occurs at the number density 
\begin{equation}
\rho_*=\biggl[\sqrt{\biggl(1+(k_2/k_1)\biggr)\ell^{-1}}\biggr]/\biggl[1+\sqrt{{\ell}\biggl(1+(k_2/k_1)\biggr)}\biggr]
\label{eq-rhostar}
\end{equation} 
The equations (\ref{eq-jpbc}) and (\ref{eq-rhostar}) reduce, respectively, 
to the equations (58) and (63) of ref.\cite{gccr09}.

%%%%%%%%%%%%%%%%%%%%%%%%%%%%%%%%%%%%%%%%%%%%%%%%%%%%%%%%%%%%%%%
\section{Nature of polysomes: non-equilibrium phase diagram of ribosome traffic}
%%%%%%%%%%%%%%%%%%%%%%%%%%%%%%%%%%%%%%%%%%%%%%%%%%%%%%%%%%%%%%%

In this section we impose open boundary conditions (OBC) which is more 
realistic for modeling translation. The entry of the ribosomes at one 
open end captures the initiation of translation while the exit of the 
ribosomes from the other open end mimics termination of translation.  

%%%%%%%%%%%%%%%%%%%%%%%%%%%%%%%%%%%%%%%%%%%%%%%%%%%%%%%%%%%%%%%
\subsection{Phase diagram}
%%%%%%%%%%%%%%%%%%%%%%%%%%%%%%%%%%%%%%%%%%%%%%%%%%%%%%%%%%%%%%%

In a multi-dimensional abstract space spanned by some of the crucial 
model parameters, we identify the distinct regions characterized by 
their distinctive properties which we describe below. The resulting 
diagram is referred to as a ``phase'' diagram although the ``phases'' 
are not in thermodynamic equilibrium; these phases are non-equilibrium 
steady states of the system. The theoretical prediction we make in 
this section can be tested by using the technique of polysome profile 
\cite{arava03,mikamo05}.

Our calculations here are based on the extremum current hypothesis (ECH) 
\cite{krug91,popkov2,hager1,hager2}
which relates the flux ${\cal J}$, under OBC, to the flux $J_{PBC}(\rho)$ 
of the same system, under PBC. We apply the ECH to our model in the 
same way in which it was used earlier for the simpler versions of 
our model \cite{basu07,gccr09}. 
We assume that the entrance and exit points of the track (i.e., the 
start and stop codons) are connected to two infinite particle 
reservoirs where the respective number densities are $\rho_-$ and 
$\rho_+$, respectively. According to ECH, for systems with a 
single {\it maximum} in the $J_{PBC}(\rho)$ function, 
\begin{eqnarray}
{\cal J}  = \mbox{max} ~~ J_{PBC}(\rho) & \mbox{if}~~ \rho_{-}~ > ~\rho~ > ~\rho_{+}
\end{eqnarray}
These relations can be utilized to draw the surfaces separating the 
dynamical phases on the phase diagram. The first step in this approach 
has already been completed by calculating the expression (\ref{eq-rhostar}) 
for $\rho_{*}$. Next, we derive the expressions for $\rho_{-}$ and 
$\rho_{+}$ which would give rise to the same rates of initiation and 
termination as indicated by $\alpha$ and $\beta$, respectively.

Suppose $P^{jump}_{-} ~(\Delta t)$ is the probability that, given an 
empty site, a ribosome will hop onto it from left in the next time 
interval $\Delta t$. It is straightforward to see that 
\begin{equation}  
P^{jump}_{-} (\Delta t) = P(\underbrace{1..............1}_{\ell}|\underline{0})(P_5\omega_{h2}+P_5^* \Omega_{h2} )\times \Delta t 
\end{equation}
where $P_5$  and $P_5^*$ are given by the expressions 
\begin{equation}
P_5^{-1}=\omega_{h2}(k_1^{-1}+k_2^{-1})
\label{eq-ssp5}
\end{equation}
and
\begin{equation}
P_5^* = (\Omega_p/\omega_p)P_5,
\label{eq-ssp5star}
\end{equation}
respectively, and as discussed in ref.\cite{gccr09}, 
\begin{equation}
P(\underbrace{1..............1}_{\ell}|\underline{0})=\rho/(1+\rho-\rho\ell)
\end{equation} 
Note that the solutions (\ref{eq-ssp5}) and (\ref{eq-ssp5star}) have been 
obtained using the normalization condition 
\begin{equation}
\sum_{\mu=1}^{5} P_{\mu} + P_{4}^{*} + P_{5}^{*} = 1
\label{eq-norm2}
\end{equation}
for the reservoir.

Now $\rho_-$ is the solution of the equation $\alpha=P^{jump}_{-}$. Hence, 
\begin{equation}
\rho_-=\alpha\biggl(1+(k_2/k_1)\biggr)/\biggl[(\ell-1)\biggl(1+(k_2/k_1)\biggr)\alpha +P_{k2}\biggl(1+(\Omega_p/\omega_p)\biggr)\biggr]
\label{eq-rhominus}
\end{equation}
with 
\begin{equation}
P_{k2}=k_2 ~(\Delta t)  
\label{eq-pk2}
\end{equation}
relating the probability $P_{k2}$ with the rate constant $k_2$ where $k_2$ 
is given by (\ref{eq-k2}).  Similarly, 
\begin{equation}
P^{jump}_{+} ~(\Delta t) = P(\overbrace{\underline{1..............1}}^{\ell}|~0)(P_5 \omega_{h2}+P_5^* \Omega_{h2})\Delta t
\end{equation}
where 
\begin{equation}
P(\overbrace{\underline{1..............1}}^{\ell}|~0)=(1-\rho\ell)/(1+\rho-\rho\ell)
\end{equation}
The unknown density $\rho_{+}$ is the solution of the equation 
$\beta=P^{jump}_{+}$; hence,
\begin{equation}
\rho_+ = \biggl[\beta\biggl(1+(k_2/k_1)\biggr)-P_{k2}\biggl(1+(\Omega_p/\omega_p)\biggr)\biggr]/\biggl[\beta(\ell-1)\biggl(1+(k_2/k_1)\biggr)-\ell P_{k2}\biggl(1+(\Omega_p/\omega_p)\biggr)\biggr]
\label{eq-rhoplus}
\end{equation}
Note that the equations (\ref{eq-rhominus}) and (\ref{eq-rhoplus}) 
reduce, respectively, to the equations (67) and (70) of ref.\cite{gccr09} 
in the appropriate limit. 

%%%%%%%%%%%%%%%%%%%%%%%%%%%%%%%%%%%%%%%%%%%%%%%%%%%%%%%%%%%%%%%%%%%
\subsubsection{Surface separating LD and MC phases }
%%%%%%%%%%%%%%%%%%%%%%%%%%%%%%%%%%%%%%%%%%%%%%%%%%%%%%%%%%%%%%%%%%%
From MCH it follows that the surface separating the LD and MC phases 
on the phase diagram of the system is obtained from the equation  
\begin{equation}
 \rho_-=\rho_*
\end{equation}
by expressing $\rho_-$ and $\rho_*$ in terms of the rate constants 
for the elementary steps of the model kinetics. Hence, the equation 
for this surface in the phase diagram of the model is found to be 
\begin{equation}
 \alpha=\biggl[\rho_* P_{k2}\biggl(1+(\Omega_p/\omega_p)\biggr)\biggr]/\biggl[\biggl(1+(k_2/k_1)\biggr)[1-(\ell-1)\rho_*]\biggr]
\label{eq-alpha1}
\end{equation}
%%%%%%%%%%%%%%%%%%%%%%%%%%%%%%%%%%%%%%%%%%%%%%%%%%%%%%%%%%%%%%%%%%%
\subsubsection{Surface separating HD and MC phases }
%%%%%%%%%%%%%%%%%%%%%%%%%%%%%%%%%%%%%%%%%%%%%%%%%%%%%%%%%%%%%%%%%%%
Similarly, using the condition  
\begin{equation}
 \rho_+=\rho_*
\end{equation}
the equation for the surface separating the HD and MC phases in the 
phase diagram of the system is given by 
\begin{equation}
 \beta=\biggl[P_{k2}(1-\rho_*\ell)\biggl(1+(\Omega_p/\omega_p)\biggr)\biggr]/\biggl[\biggl(1+(k_2/k_1)\biggr)[1-(\ell-1)\rho_*]\biggr]
\label{eq-beta1}
\end{equation}
%%%%%%%%%%%%%%%%%%%%%%%%%%%%%%%%%%%%%%%%%%%%%%%%%%%%%%%%%%%%%%%%%%%
\subsubsection{Surface of coexistence of  HD and LD phases }
%%%%%%%%%%%%%%%%%%%%%%%%%%%%%%%%%%%%%%%%%%%%%%%%%%%%%%%%%%%%%%%%%%%
\begin{equation}
J_{PBC}(\rho_-)=J_{PBC}(\rho_+)
\end{equation}
which gives 
\begin{equation}
\alpha=\biggl[P_{k2}\biggl(1+(\Omega_p/\omega_p)\biggr)\beta\biggl(1+(k_2/k_1)\biggr)\biggr]/\biggl[P_{k2}\biggl(1+(\Omega_p/\omega_p)\biggr)\ell+\beta\biggl(1-\ell+2(k_2/k_1)-\ell(k_2/k_1)+(k_2^2/k_1^2)\biggr)\biggr]
\label{eq-alpha2}
\end{equation}
Equivalently
\begin{equation}
\beta=\biggl[\alpha \ell P_{k2}\biggl(1+(\Omega_p/\omega_p)\biggr)\biggr]/\biggl[\biggl(1+(k_2/k_1)\biggr)\biggl\{P_{k2}\biggl(1+(\Omega_p/\omega_p)\biggr)+\alpha(\ell-1)-(k_2/k_1)\alpha\biggr\}\biggr]
\label{eq-beta2}
\end{equation}
%%%%%%%%%%%%%%%%%%%%%%%%%%%%%%%%%%%%%%%%%%%%%%%%%%%%%%%%%%%%%%%
Note the equations (\ref{eq-alpha1}), (\ref{eq-beta1}), (\ref{eq-alpha2}) 
and (\ref{eq-beta2}) reduce to the expressions (72), (74), (76) and 
(77), respectively, of ref.\cite{gccr09} in the appropriate limit.

Figs.\ref{fig-abphi1} and \ref{fig-abphi2} show the 3-dimensional 
phase diagrams plotted in the $\alpha-\beta-\phi$ space from two 
different perspectives, while figs.\ref{fig-abr21} and \ref{fig-abr22} 
show the 3-dimensional phase diagrams plotted in the 
$\alpha-\beta-P_{\omega_{r2}}$ space also from two different 
perspectives. Since none of the earlier models of ribosome traffic 
capture translational fidelity and kinetic proofreading explicitly, 
these phase diagrams have not been reported ever before.

In order to compare the implications of these phase diagrams with 
that of the TASEP, we also project several two-dimensional cross 
sections of these phase diagrams onto the $\alpha-\beta$-plane 
(see figs.\ref{fig-2dphi} and \ref{fig-2dr2}). 
On the 2D projections, transition from the LD phase to the MC 
phase takes place at $\alpha = \alpha_{*}$ whereas the transition 
from the HD phase to the MC phase takes place at $\beta=\beta_{*}$. 

As is evident from these 2D phase diagrams, the curvature of the 
lines of coexistence of HD and LD phases seems to be the generic 
feature of all such models \cite{gccr09,antal00}. 
The straight line $\alpha=\beta$ on which the LD and HD phases of 
TASEP coexist is a manifestiation of the ``particle-hole'' symmetry 
in TASEP, a special property that is not shared by our model of 
ribosome traffic. 

Increasing $\phi$ shifts $\alpha_{*}$ and $\beta_{*}$ to higher 
values. Increase of $\phi = \omega_p/(\Omega_p+\omega_p)$ can 
be viewed as a result of increasing $\omega_{p}$ which, in turn, 
increases the effective rate of hopping of a ribosome from one 
codon to the next. It is well known \cite{kolo98} that higher 
values of effective hopping rate shifts $\alpha_{*}$ and $\beta_{*}$ 
to higher values. Similarly, increasing $\omega_{r2}$ decreases 
the effective hopping rate thereby shifting $\alpha_{*}$ and 
$\beta_{*}$ to smaller values.

%%%%%%%%%%%%%%%%%%%%%%%%%%%%%%%%%%%%%%%%%%%%%%%%%%%%%%%%%%%%%%
\subsection{Effects of recycling on the phase diagram}
%%%%%%%%%%%%%%%%%%%%%%%%%%%%%%%%%%%%%%%%%%%%%%%%%%%%%%%%%%%%%%

Recycling of ribosomes can be captured by our model by replacing 
the constant initiation rate $\alpha$ by an {\it effective} 
initiation rate $\alpha_{eff}$. Since the availability of ribosomes 
for initiation is proportional to the flux of ribosomes exiting 
from the stop codon, we postulate that 
\begin{equation}
\alpha_{eff}=\alpha+qJ(\alpha_{eff},\beta,\{\omega_i\})
\label{eq-alpha}
\end{equation}
where the coefficient $q$ depends on the relative separation between the  
two ends of the $mRNA$ transcript as well as on the diffusion constant 
of the ribosome subunits in the solution.\cite{chou03}.
Note that the prescription (\ref{eq-alpha}) for recycling is similar 
in spirit, but not identical quantitatively, to the prescription used 
by Gilchrist and Wagner \cite{gilchrist06} because the latter model 
does not capture steric exclusion among the ribosomes. 

On simple physical grounds, the effect of (\ref{eq-alpha}) on the phase 
diagram is expected to be non-trivial. Suppose, $\alpha$ is varied keeping 
$\beta$ fixed. At a very small value of $\alpha$ ($\alpha \ll \beta$), 
the flux would be determined by $\alpha$. Starting from a very small value, 
increasing $\alpha$ initially increases $J$ which, in turn, increases 
the effective initiation rate $\alpha_{eff}$. However, beyond a certain 
value of $\alpha$, $\alpha_{eff}$ is no longer rate limiting and  
the flux becomes independent of $\alpha$.

Exploiting the relation between our model and $\ell$-TASEP, we plot 
the phase diagram for the extended model that captures recycling of 
ribosomes through $\alpha_{eff}$. We use the results reported in 
refs.\cite{shaw03,zia11} for $\ell$-$TASEP$ and get 
\begin{equation}
 J=[\alpha_{eff}(1-\alpha_{eff})]/[1+\alpha_{eff}(\ell-1)]
\end{equation}
\begin{equation}
\alpha_{eff}=\alpha+q\biggl([\alpha_{eff}(1-\alpha_{eff})]/[1+\alpha_{eff}(\ell-1)]\biggr)
\end{equation}
which gives
\begin{equation}
\alpha_{eff}=[\alpha(\ell-1)+q-1+\sqrt{(\alpha(\ell-1)+q-1)^2+4(\ell-1+q)\alpha}]/[2(\ell-1+q)]
\end{equation}
So at LD and HD interface
\begin{equation}
\beta=\alpha_{eff}
\end{equation}
at LD and MC boundary
\begin{equation}
\alpha_{eff}=1/(1+\sqrt{\ell}) 
\end{equation}
and at HD and MC boundary 
\begin{equation}
\beta=1/(1+\sqrt{\ell}) 
\end{equation}
The resulting 2-d phase diagram in the $\alpha-\beta$ plane is plotted 
in fig.\ref{fig-recycle}. Increase of the recycling factor $q$ shifts 
$\alpha_{*}$ to a smaller value while $\beta_{*}$ remains unchanged. 
This trend of variation is consistent with the fact that recycling 
affects only the initiation rate without influencing the rate of 
termination. This is consistent with the trends of variation of the 
coverage density and flux of ribosomes that we observed in our 
computer simulations which we descibe below.

%%%%%%%%%%%%%%%%%%%%%%%%%%%%%%%%%%%%%%%%%%%%%%%%%%%%%%%%%%
\subsubsection{Variations of flux and coverage density with the extent of recycling}
%%%%%%%%%%%%%%%%%%%%%%%%%%%%%%%%%%%%%%%%%%%%%%%%%%%%%%%%%%

In order to demonstrate the effects of recycling in a form that would 
be closer to one's physical intuition, we now show the variation of 
the average flux and the average coverage density of the ribosomes 
with the parameter $q$ which is a measure of the extent of recycling. 
For this purpose, we carried out computer simulations of our model 
for several different values of $q$. All the data reported here were 
obtained for $L = 1000$ and $\ell = 10$.  
Since the linear size of a ribosome is measured here in the units of 
codons, ${\ell} = 10$ is a realistic choice because in recent 
experiments \cite{ingolia10,guo10} it has been observed that each 
ribosome covers about 30 nucleotides on the mRNA track. 
Typical values of some of the rate constants have been
reported in the literature \cite{thompson80,harrington93}. 
For those rate constants whose numerical values are not available in 
the literature, we have assumed some reasonable values based on physical 
intuition. However, our conclusions are not sensitive to the precise 
numerical values of the rate constants.

The rate constants which we used for the simulations are 
$\omega_a$=25$s^{-1}$, $\omega_{h1}$=25$s^{-1}$,$\omega_{h2}$=25$s^{-1}$,$\omega_p$=25$s^{-1}$,$\omega_{bf}$=25$s^{-1}$,$\omega_{br}$=25$s^{-1}$,$\omega_{r1}=$10$s^{-1}$,$\omega_{r2}=$10$s^{-1}$,$\Omega_p$=5$s^{-1}$,$\Omega_{bf}$=5$s^{-1}$,$\omega_{br}$=5$s^{-1}$,$\omega_{h2}$=5$s^{-1}$\\
All the rate constants were converted to dimensionless transition 
probabilities using the formula $P_{\omega}=1-exp(-\omega*(\Delta t))$ 
where time step $\Delta t=0.005$s is used for all the simulation runs. 

At first sight, it may appear that there is some ambiguity in the 
defnition of $\alpha_{eff}$: what value of $J$ should be used in 
(\ref{eq-alpha})? In order to avoid ambiguity, we average the 
spatially-averaged flux further over the time period elapsed since 
the last entry of a ribosome (i.e., the initiation of translation 
by the ribosome closest to the start codon). This ``doubly-averaged'' 
value of flux $J$ is used in (\ref{eq-alpha}) to compute $\alpha_{eff}$.

In fig(\ref{fig-funda}) we plot the flux and the coverage density of 
the ribosomes as functions of $q$. Starting from a vanishingly small 
value, as $q$ is increased, both the flux and coverage density increase. 
However, beyond a limiting value, both become practically independent 
of $q$; this trend of variation is caused by a transition from the LD 
phase to either the HD phase or to the MC phase, depending on the 
values of the set of other parameters.

`%%%%%%%%%%%%%%%%%%%%%%%%%%%%%%%%%%%%%%%%%%%%%%%%%%%%%%%%%%%%%%%%%%%%
\subsection{Experimental tests of the phase diagram with polysome profile}
%%%%%%%%%%%%%%%%%%%%%%%%%%%%%%%%%%%%%%%%%%%%%%%%%%%%%%%%%%%%%%%%%%%%

The three different phases are characterized by three different densities;  
the expressions for these densities have been derived above. Therefore, our 
theoretical predictions can be tested by measuring the average densities. 
For this purpose, {\it polysome profiling} \cite{arava03,mikamo05} would 
be adequate. However, all the analytical calculations for the phase diagram 
have been carried out for sequence-homogeneous mRNA strands. Therefore, 
a poly-U strand of mRNA, with appropriate start and stop codons 
\cite{sharma11}, should be used in the experiment.

`%%%%%%%%%%%%%%%%%%%%%%%%%%%%%%%%%%%%%%%%%%%%%%%%%%%%%%%%%%%%%%%%%%%%
\section{Instantaneous spatial distribution of ribosomes: ribosome profile}
%%%%%%%%%%%%%%%%%%%%%%%%%%%%%%%%%%%%%%%%%%%%%%%%%%%%%%%%%%%%%%%%%%%%

In all the sections above we have calculated quantitative 
characteristics which do not require information on the spatial 
distributions of the ribosomes on the mRNA transcript. In this 
section we explore some other quantitative features of ribosome 
traffic which deal with the spatial distributions of the ribosomes. 
Our theoretical predictions on the spatial distributions of the 
ribosomes can be tested with the ribosome profiling technique 
\cite{guo10,ingolia10}.

`%%%%%%%%%%%%%%%%%%%%%%%%%%%%%%%%%%%%%%%%%%%%%%%%%%%%%%%%%%%%%%%%%%%%
\subsection{Distance-headway distribution}
%%%%%%%%%%%%%%%%%%%%%%%%%%%%%%%%%%%%%%%%%%%%%%%%%%%%%%%%%%%%%%%%%%%%

The distance-headway (DH) is defined as the {\it spatial} separation 
between two successive ribosomes on the same mRNA transcript. 
At any given instant of time, the magnitude of the DH fluctuates 
from one pair of ribosomes to another, the instantaneous spatial 
distribution of the ribosmes is characterized by the corresponding 
distribution of the DHs. The DH distribution is used extensively 
for quantitative characterization of macroscopic vehicular traffic 
\cite{css00,scn10}. In this subsection we calculate the DH distribution 
for our kinetic model of ribosome traffic. Since ribosome profiling 
\cite{ingolia10,guo10}  provides the exact positions of the ribosomes 
at the instant when translation was stopped, DH distribution can be 
extracted by repeatiting this profiling sufficiently large number of 
times. 

Our system may be viewed as one that consists of $M$ identical rods, 
each of length $\ell$, distributed over a lattice of $L$ sites. 
First, assuming a ring-like mRNA track we get the DH distribution 
for the corresponding number density $\rho$ which, because of the PBC, 
does not fluctuate. In this case, the expression for the DH distribution 
is given by \cite{shaw03} 
\begin{equation}
P_{dh}(m,\rho)=(\rho/\rho_s)(\rho_h/\rho_s)^m 
\end{equation}
where $\rho_h = 1-\rho\ell$ is the density of holes and
$\rho_s=\rho+\rho_{h}=1+\rho-\rho\ell$. Hence, 
\begin{equation}
P_{dh}(m,\rho)=[\rho(1-\rho\ell)^{m}]/[(1+\rho-\rho\ell)^{m+1}]
\label{eq-DHfinal}
\end{equation}

The number density of the ribosomes for the real system under OBC is a 
fluctuating quantity. But, the mean density deep inside the bulk (around 
the central region of the lattice) can be extracted numerically from 
computer simulations. Substituting the numerically estimated density of 
the ribosomes under OBC into the expression (\ref{eq-DHfinal}) we get the DH 
distribution under OBC. This distribution is plotted in fig.\ref{fig-DH}. 
The straight lines on the semi-log plot reflects the geometric nature of 
the distribution (discrete analog of the exponential distribution). 
In order to test the validity of the approximate scheme used above to 
derive the DH distribution by a combination of analytical and numerical 
arguments, we have also computed the DH distribution directly by computer 
simulation; the simulation data are also plotted in fig.\ref{fig-DH}. 
The theoretically derived lines are in reasonaly good agreemwnt with 
the DH distribution obtained by computer simulation.

`%%%%%%%%%%%%%%%%%%%%%%%%%%%%%%%%%%%%%%%%%%%%%%%%%%%%%%%%%%%%%%%%%%%%
\subsection{Influence of slow codons on the density profile and flux}
%%%%%%%%%%%%%%%%%%%%%%%%%%%%%%%%%%%%%%%%%%%%%%%%%%%%%%%%%%%%%%%%%%%%

It is well known that translation of some codons take place at a 
very slow rate; these are often referred to as ``hungry'' codon. 
However, we'll use the term ``slow'' to refer to the all those 
codons which get translated at a much slower rate than other codons. 
In this subsection we explore the effects of bottlenecks created 
by such slow codons against the forward movements of ribosomes. 
In particular, we investigate the effects of slow codons on the 
average density profile and flux of ribosomes in ribosome traffic. 

In the model that we simulated for this purpose, a $mRNA$ transcript 
consists of only two different types of codons. 
In the computer simulations of our model we assign ten times smaller 
numerical value to $\omega_a$ for a slow codon compared to that 
of a normal codon. Simultaneously, the numerical value of $\omega_{r1}$ 
assigned to a slow codon is ten times larger than that of a normal 
codon. Since in this particular study we are interested mainly in 
the effects of bottlenecks, we ignore the possibility of misincorporation 
by erasing the branched pathway putting $\Omega_p = 0 = \Omega_{h2}$. 

In the first set of simulations, we put four slow codons at the 
center of the stretch of mRNA between the start and the stop codon.
Such a single extended bottleneck leads to a ``phase-segregated'' 
profile where on one side of the bottleneck the average density 
is much higher than that on the other side (see fig\ref{scfigure11}).  
Profiles are plotted for different values of $\omega_{r2}$.  
Higher value of $\omega_{r2}$ reduce the effective hopping rate 
for both normal as well as rare codons. But, the effective hopping 
rate for slow codon decrease more because of higher value of  
$\omega_{r1}/\omega_{h1}$. Thus, the higher is the value 
of $\omega_{r2}$, the larger is the difference between the effective 
rates of hopping from normal and slow codons. This, in turn, leads 
to the larger jump discontinuity of the density across the bottleneck 
at a larger value of $\omega_{r2}$. 

In the second set of simulations, the slow codons were not clustered 
together. Instead, four equispaced slow codons were placed at the 
sites $200, 400, 500$, and $800$ on a lattice of total length 
$L = 1000$.The average density profiles for this case are plotted 
in fig.\ref{scfigure12} for two different values of $\omega_{r2}$.
Both the profiles exhibit discontinuous jumps in the coverage 
density; the position of each minimum in the coverage density coincides 
with the location of a slow codon. Moreover, periodic oscillations are 
also observed in the vicinity of the the rare codon where periodicity  
is $\ell$. Similar results were obtained earlier in TASEP-type 
models of ribosome traffic \cite{zia11}; however, unlike our data, 
shown in fig.\ref{scfigure12}, the effects of kinetic proofreading 
and futile cycles could not be addressed by the model of ref.\cite{zia11}.

In order to emphasize the effect of clustering of slow codons on 
the overall rate of protein synthesis we plot flux as a function 
of $\omega_{r2}$ in fig\ref{scfigure13} for the two different 
conditions discussed above. The upper curve corresponds to the 
setup where four slow codons are placed equidistant on lattice. 
The lower curve corresponds to the setup where all the four 
slow codons are placed clustered together at the center of the system.
The data clearly show that without increasing the number of slow 
codons the rate of synthesis of proteins can be reduced drastically 
by clustering the slow codons into a single bottleneck.

`%%%%%%%%%%%%%%%%%%%%%%%%%%%%%%%%%%%%%%%%%%%%%%%%%%%%%%%%%%%%%%%%%%%%
\subsubsection{Probing spatial distribution of ribosomes}
%%%%%%%%%%%%%%%%%%%%%%%%%%%%%%%%%%%%%%%%%%%%%%%%%%%%%%%%%%%%%%%%%%%%

Ribosome profiling technique \cite{ingolia10,guo10} is ideally suited to 
probe the instantaneous spatial distribution of ribosomes on the 
same mRNA transcript.  But, our model does not take into account 
the variation of rate constants arising from sequence inhomogeneity 
of the mRNA transcript. Therefore, at first sight, it may appear 
that a homogeneous sequence (e.g., a poly-U) would be most appropriate 
transcript for testing our theoretical prediction. However, the 
technique of ribosome profiling \cite{ingolia10,guo10} 
cannot locate the exact positions of the ribosomes on a sequence-
homogeneous mRNA transcript. Therefore, we suggest that the 
experiment should be performed with a special-type of 
sequence-inhomogeneous mRNA transcript where, because of the intrinsic 
degeneracy of the genetic code, all the codons are {\it synonymous}, 
i.e., correspond to the same amino acid. Only the cognate tRNA 
molecules carrying the correct amino acid are to be supplied to 
the solution. We do not expect significant codon-to-codon variation 
of the rate constants in this case. For studying the effects of 
translational fidelity and proofreading, tRNA molecules carrying a 
non-cognate species of amino-acids should also be supplied.

In case it turns out to be difficult to extract the exact positions 
of the ribosomes from ribosome profiling of such a mRNA strand 
where the degenerate codons are distributed randomly, we suggest 
an alternative strategy. Recall that six synonymous codons 
correspond to the same amino acid Arg; similarly there is 
six-fold degeneracy also for the amino acids Leu and ser 
(see fig.\ref{fig-sequence}(a)). In principle, one can synthesize 
an artificial mRNA transcript of the type shown in 
fig.\ref{fig-sequence}(b) using six synonymous codons from any 
of the three possible sets shown in fig.\ref{fig-sequence}(a).  
A mRNA transcript with such a non-random inhomogeneous codon sequence 
allows unambiguous identification of the positions of the ribosomes 
on it while using the ribosome profiling technique \cite{ingolia10}. 

%%%%%%%%%%%%%%%%%%%%%%%%%%%%%%%%%%%%%%%%%%%%%%%%%%%%%%%%%%%%
\section{Summary and conclusion} 
%%%%%%%%%%%%%%%%%%%%%%%%%%%%%%%%%%%%%%%%%%%%%%%%%%%%%%%%%%%%

In this paper we have developed a theoretical framework that 
captures several key features of translation as well as the 
spatio-temporal organization of polysomes. First, the selection 
of aa-tRNA in our model is a two-stage process; the second 
stage captures kinetic proofreading. Second, our model allows 
occasional translational error and we calculate several 
quantities as functions of the translational fidelity. 
We have also incorporated some of the other features of the 
mechano-chemical cycle of ribosomes in the elongation stage 
which, to our knowledge, have not been incorporated in any 
earlier model. On the basis of our hypothesis for capturing 
the effects of ribosome recycling, we have predicted the 
effects of recycling on the spatial profile of the ribosomes 
as well as on the rate of protein synthesis. 

Here we have also investigated the spatial organization 
of ribosomes in polysomes in terms of the distance-headway 
distribution; it is a quantity that is used extensively to 
characterize crowding in vehicular traffic. We have also 
identified the parameter regimes which display distinct 
characters of polysomes and the corresponding rates of 
protein production in our model system. Finally, we have 
also demonstrated the effects of sequence inhomogeneity of the 
mRNA transcript, particularly, that of the clustering of slow 
codons.  

We hope this work will inspire experimental investigation 
for measuring new quantities. It is possible to test 
some of our new predictions using polysome profile techniques.  
But, more interesting results on spatial distributions 
of the ribosomes on the mRNA transcript would require 
ribosome profiling \cite{ingolia10,guo10}.

%%%%%%%%%%%%%%%%%%%%%%%%%%%%%%%%%%%%%%%%%%%%%%%%%%%%%%%%%%%%
\noindent{\bf Acknowledgements} 
%%%%%%%%%%%%%%%%%%%%%%%%%%%%%%%%%%%%%%%%%%%%%%%%%%%%%%%%%%%%

This work is supported by IIT Kanpur through the Dr. Jag Mohan 
Chair professorship to one of the authors (DC). DC also thanks 
the visitors program of the Max-Planck Institute for Physics 
of Complex Systems for hospitality in Dresden during the 
preparation of this manuscript. 

%%%%%%%%%%%%%%%%%%%%%%%%%%%%%%%%%%%%%%%%%%%%%%%%%%%%%%%%%%%%
\section{References}
%%%%%%%%%%%%%%%%%%%%%%%%%%%%%%%%%%%%%%%%%%%%%%%%%%%%%%%%%%%%

%%%%%%%%%%%%%%%%%%%%%%%%%%%%%%%%%%%%%%%%%%%%%%%%%%%%%%%%%%%%%%%%

\newpage 
%%%%%%%%%%%%%%%%%%%%%%%%%%%%%%%%%%%%%%%%%%%%%%%%%%%%%%%%%%%%%%%%
\centerline{\bf Figure Captions} 
%%%%%%%%%%%%%%%%%%%%%%%%%%%%%%%%%%%%%%%%%%%%%%%%%%%%%%%%%%%%%%%%

\noindent{\bf Fig.1}: Detailed mechanochemical cycle of Ribosome on its track. 
The integer indices $...,i-1,i,i+1...$ label the codons on the mRNA transcript.  
Although the same set of transitions are allowed from each codon, only those 
from (and to) the codon $i$ are shown explicitly.\\

\noindent{\bf Fig.2}: Phase diagram of ribosome traffic model in the
3-dimensional space spanned by $\alpha$,$\beta$ and $\phi$.\\

\begin{figure}
\begin{center}
\includegraphics[angle=0,width=0.75\columnwidth]{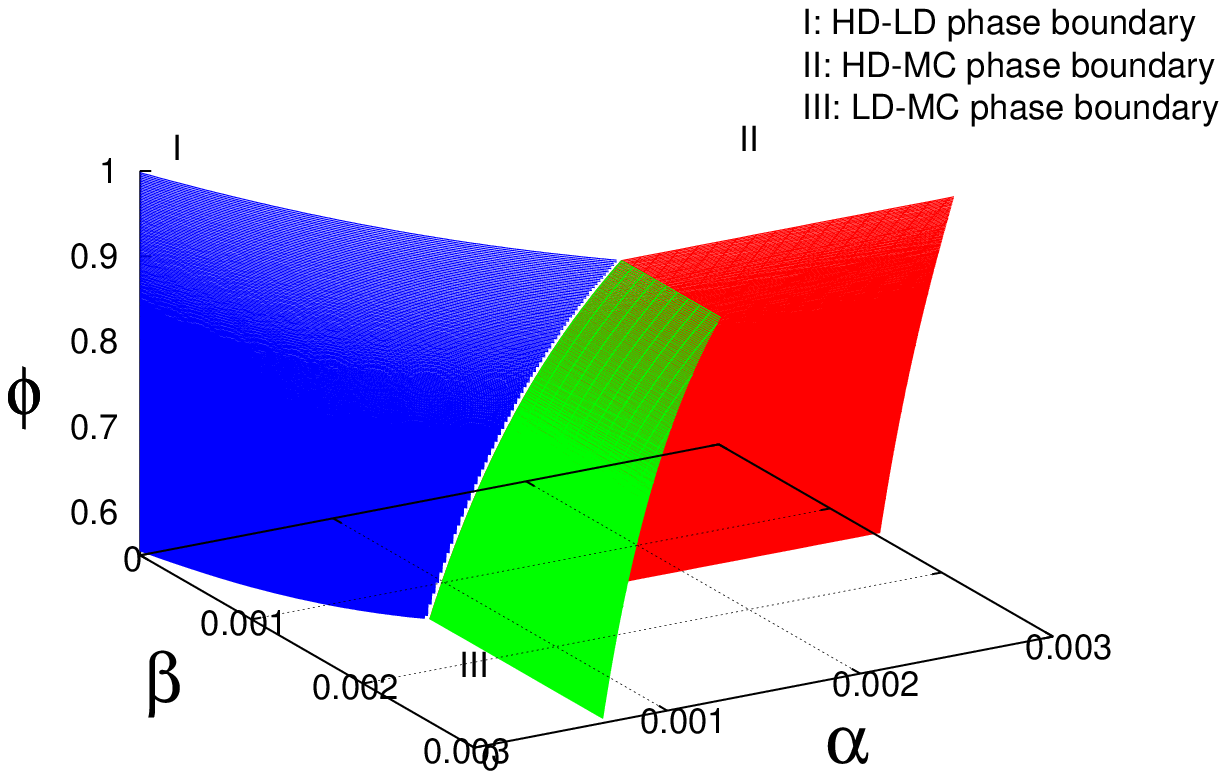}
\end{center}
\caption{} 
\label{fig-abphi1}
\end{figure}

\noindent{\bf Fig.3}: Same data as in fig.\ref{fig-abphi1}, except that
plotted from a different perspective. \\

\noindent{\bf Fig.4}: Phase diagram of ribosome traffic model in the
3-dimensional space spanned by $\alpha$,$\beta$ and $P_{\omega_{r2}}$. \\

\noindent{\bf Fig.5}: Same data as in fig.\ref{fig-abr21}, except that
plotted from a different perspective.\\

\begin{figure}
\begin{center}
\includegraphics[angle=0,width=0.75\columnwidth]{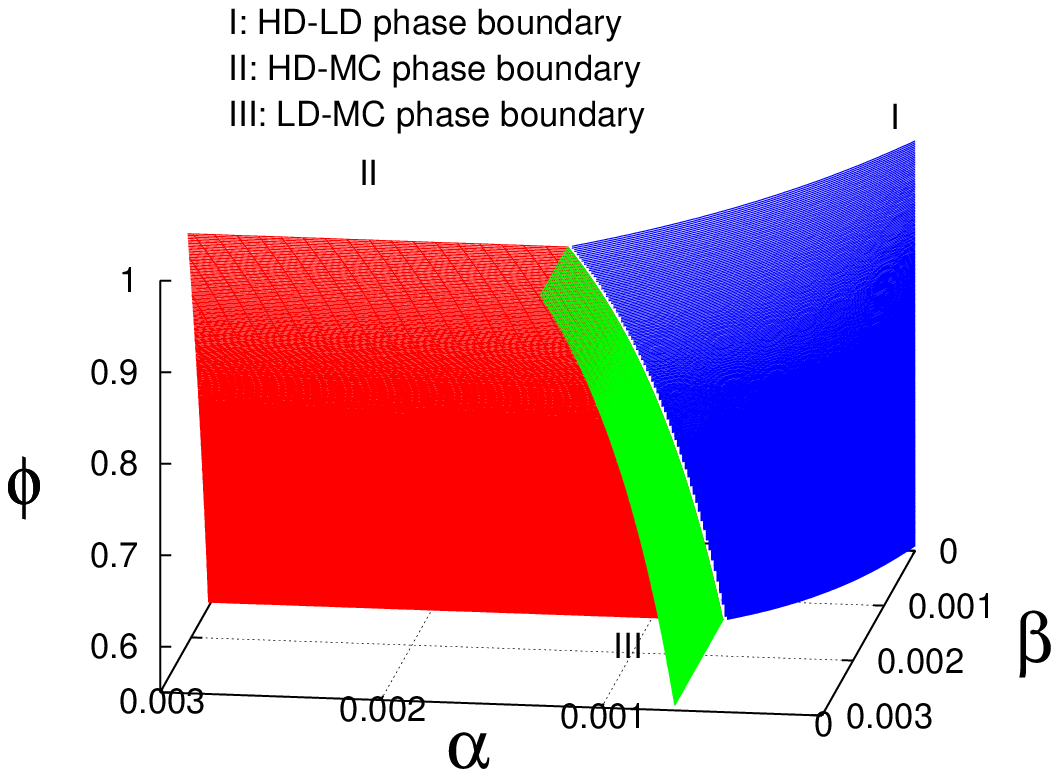}
\end{center}
\caption{} 
\label{fig-abphi2}
\end{figure}

\begin{figure}
\begin{center}
\includegraphics[angle=0,width=0.75\columnwidth]{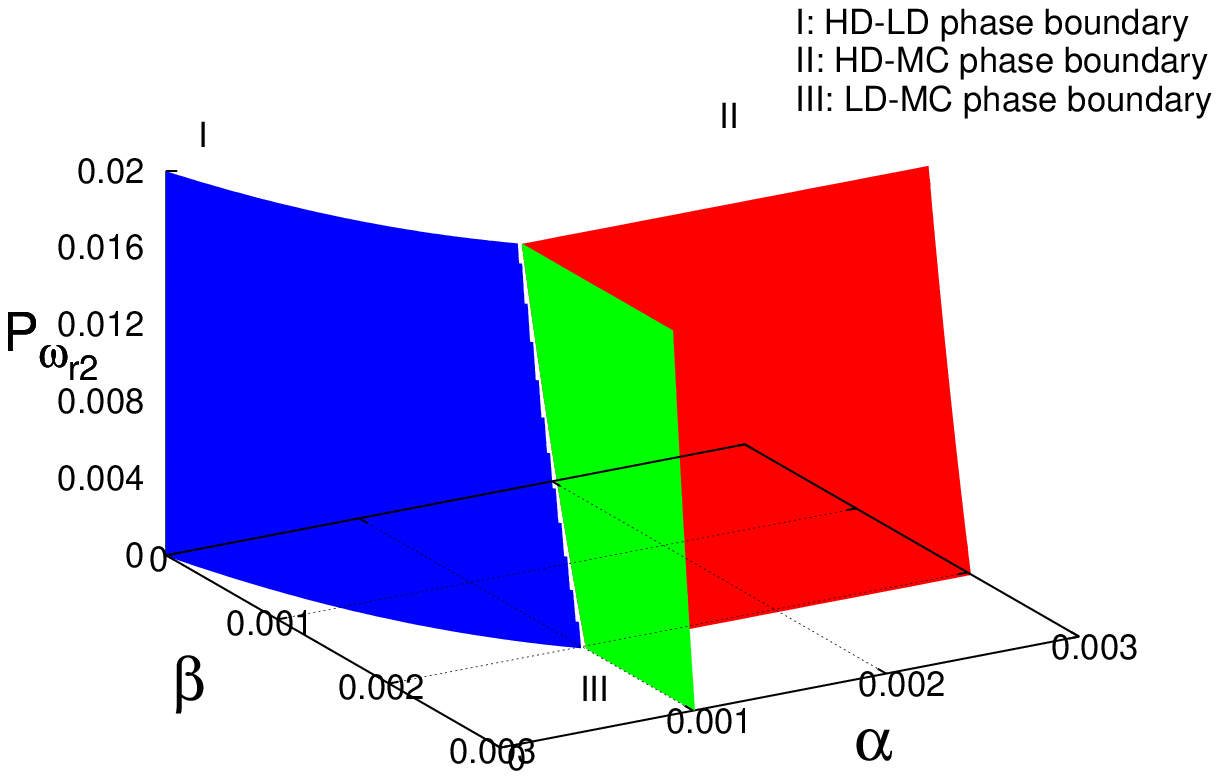}
\end{center}
\caption{} 
\label{fig-abr21}
\end{figure}

\noindent{\bf Fig.6}: Projections of several two-dimensional cross sections,
of the three-dimensional phase diagram, plotted in figs.\ref{fig-abphi1}
and \ref{fig-abphi2}, onto the $\alpha-\beta$ plane. Each
cross section corresponds to a fixed value of $\phi$. \\

\begin{figure}
\begin{center}
\includegraphics[angle=0,width=0.75\columnwidth]{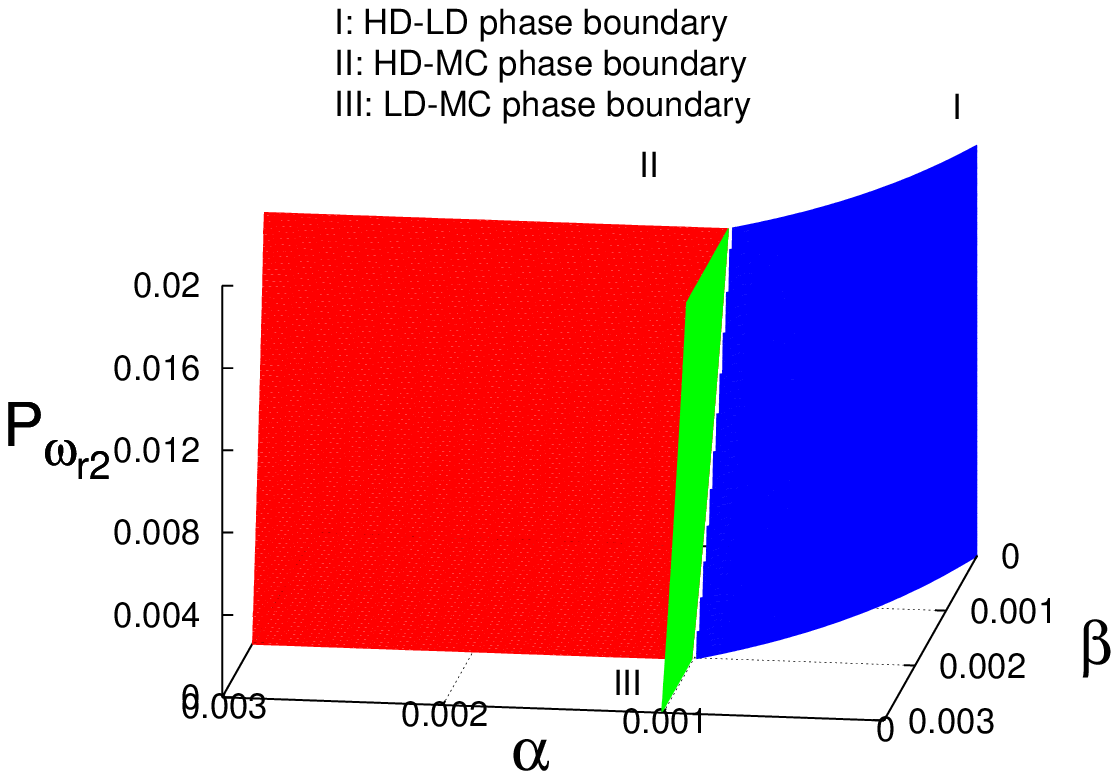}
\end{center}
\caption{} 
\label{fig-abr22}
\end{figure}

\begin{figure}
\begin{center}
\includegraphics[angle=-90,width=0.75\columnwidth]{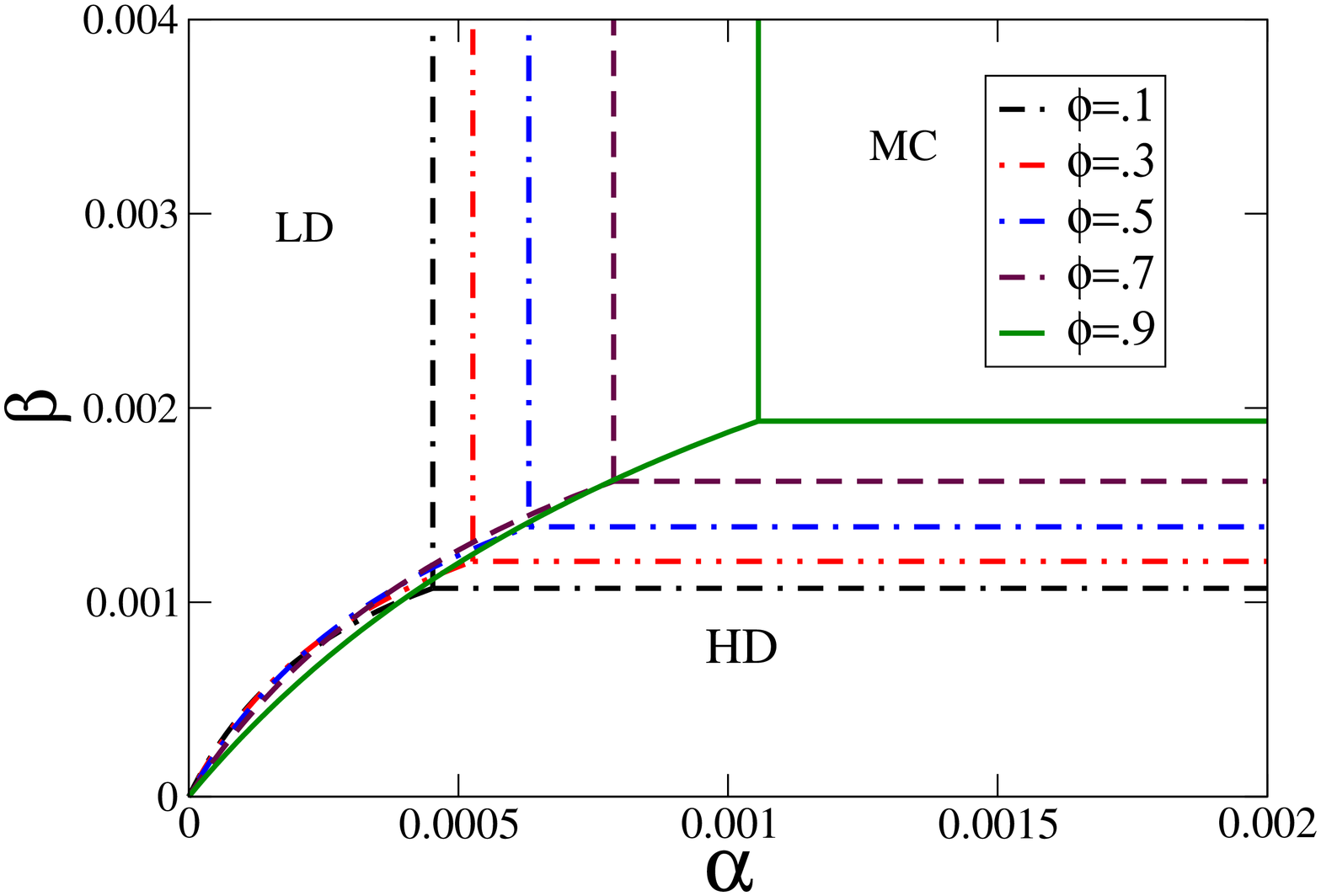}
\end{center}
\caption{} 
\label{fig-2dphi}
\end{figure}

\begin{figure}
\begin{center}
\includegraphics[angle=-90,width=0.75\columnwidth]{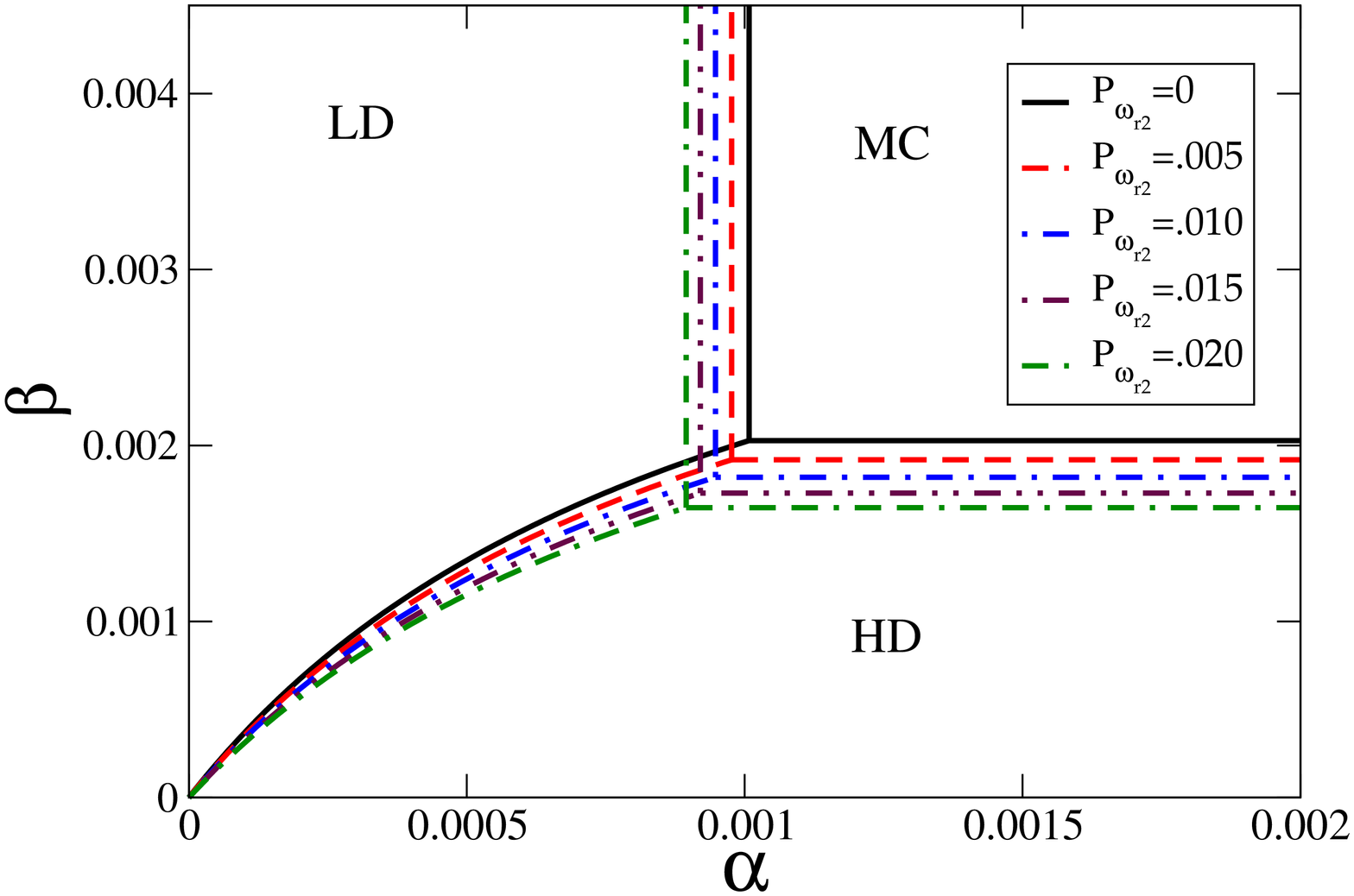}
\end{center}
\caption{} 
\label{fig-2dr2}
\end{figure}

\begin{figure}[ht]
\begin{center}
\includegraphics[angle=-90,width=0.75\columnwidth]{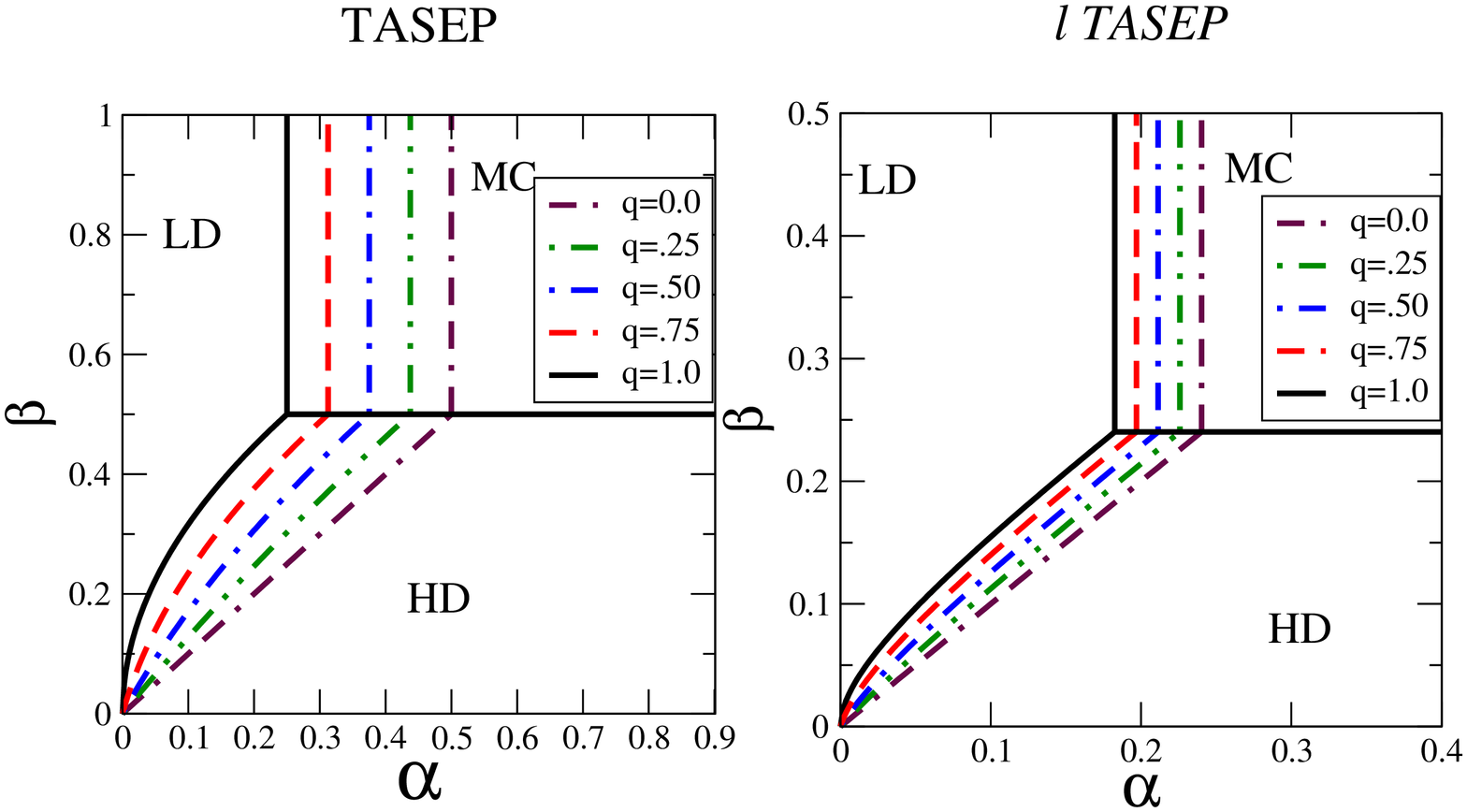}
\end{center}
\caption{} 
\label{fig-recycle}
\end{figure}

\begin{figure}[ht]
\begin{center}
\includegraphics[angle=-90,width=0.75\columnwidth]{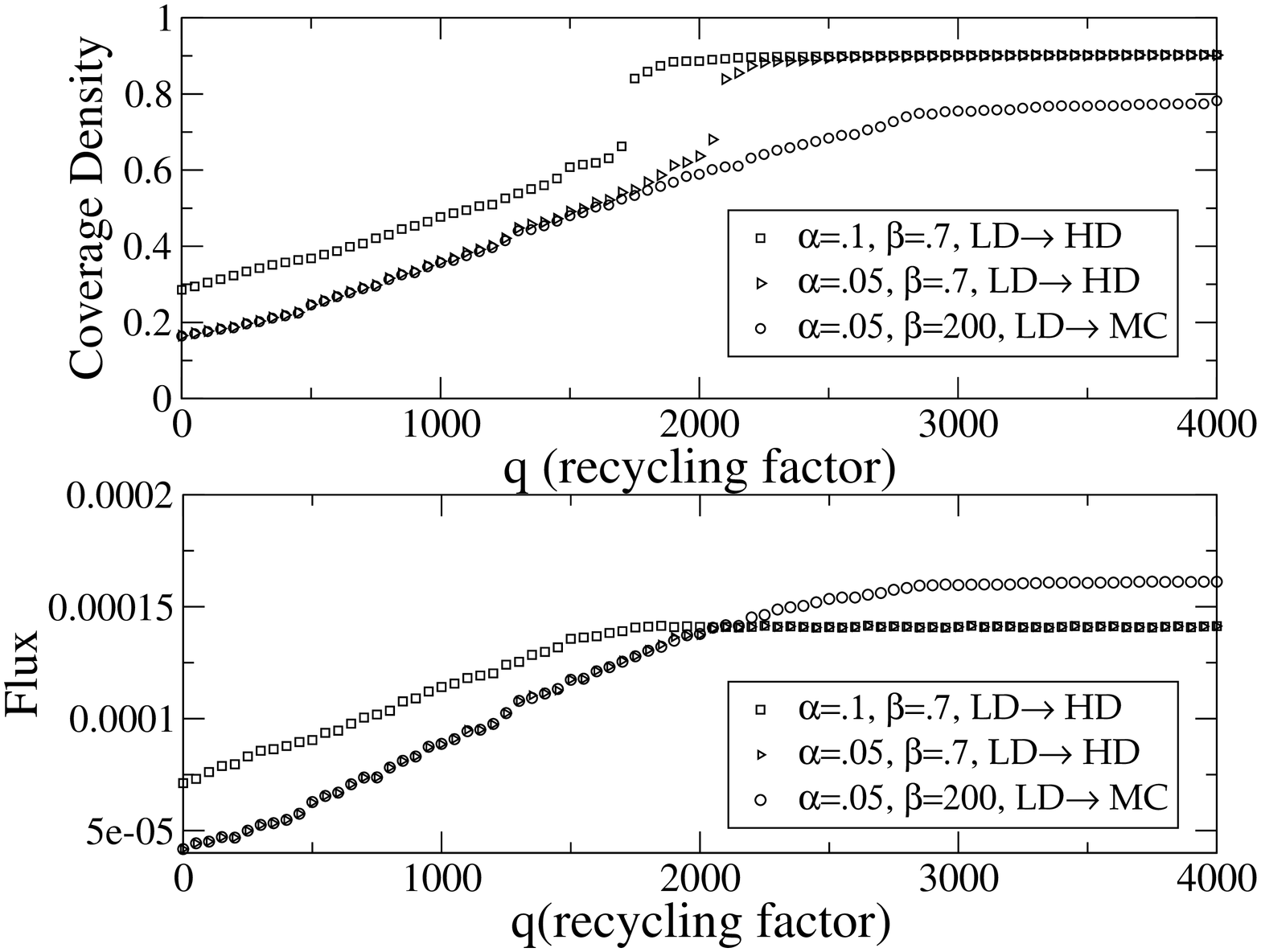}
\end{center}
\caption{} 
\label{fig-funda}
\end{figure}

\begin{figure}[ht]
\begin{center}
\includegraphics[angle=-90,width=0.75\columnwidth]{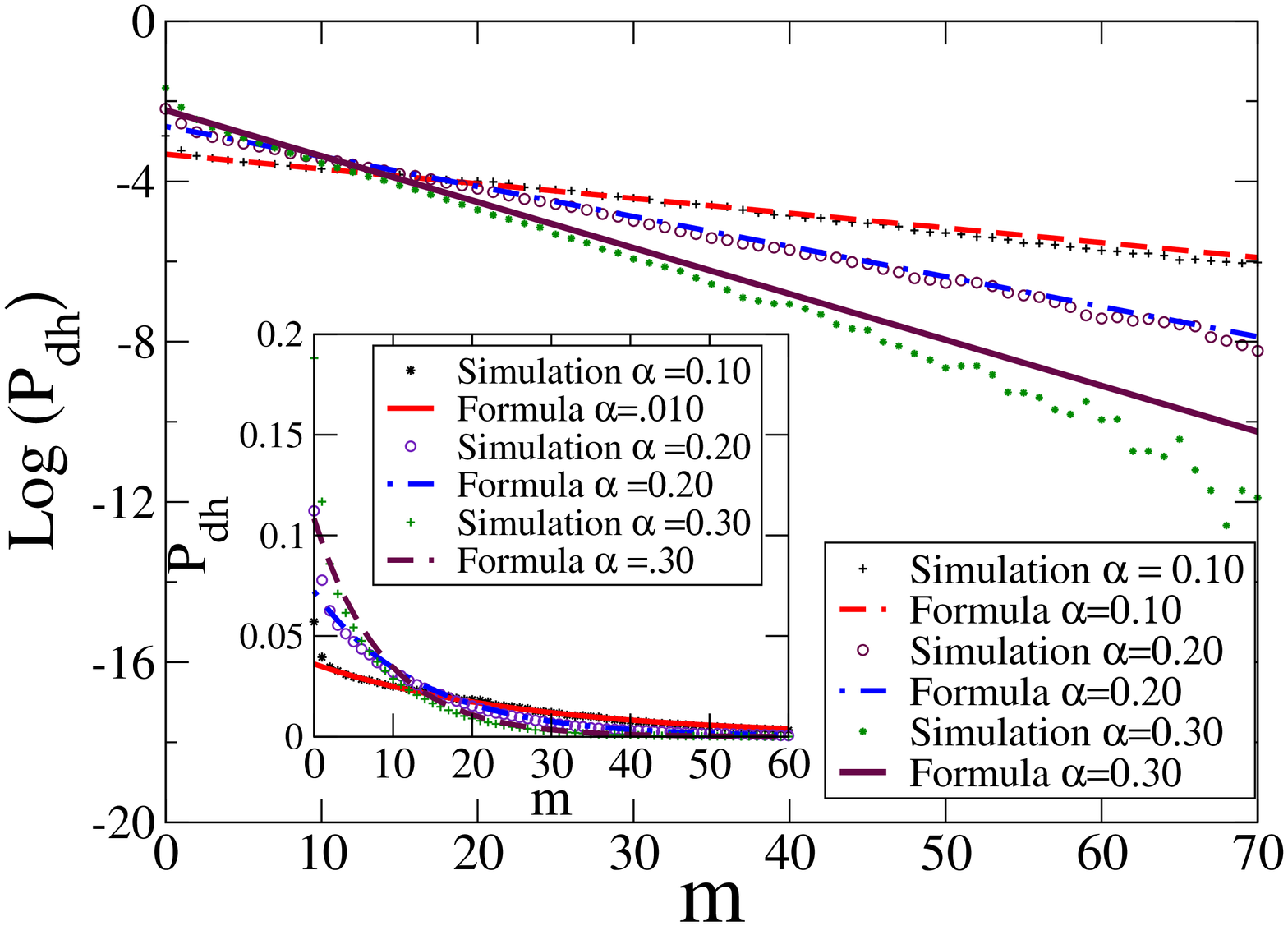}
\end{center}
\caption{} 
\label{fig-DH}
\end{figure}

\begin{figure}[ht]
\begin{center}
\includegraphics[angle=-90,width=0.75\columnwidth]{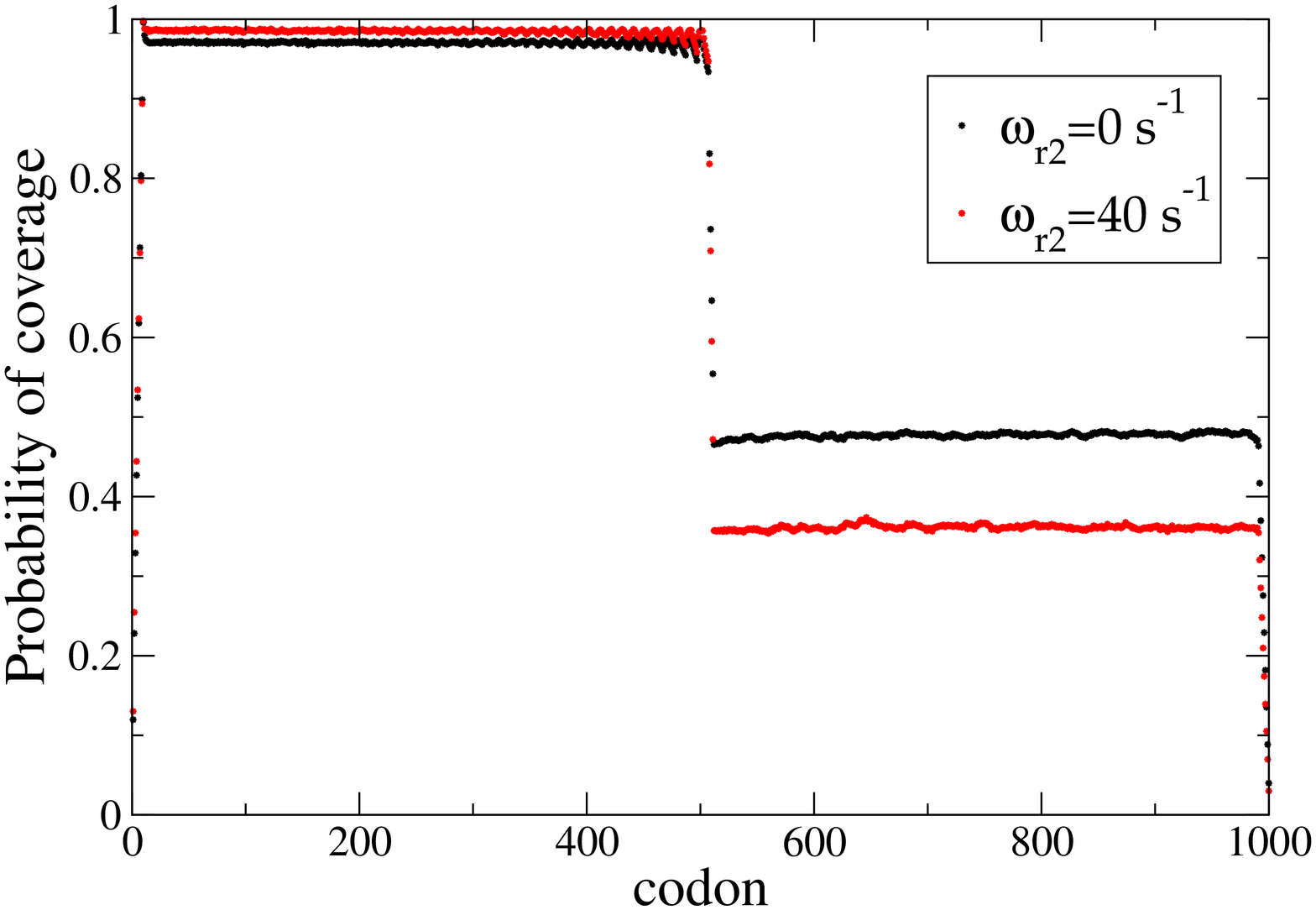}
\end{center}
\caption{} 
\label{scfigure11}
\end{figure}

\begin{figure}[ht]
\begin{center}
\includegraphics[angle=-90,width=0.75\columnwidth]{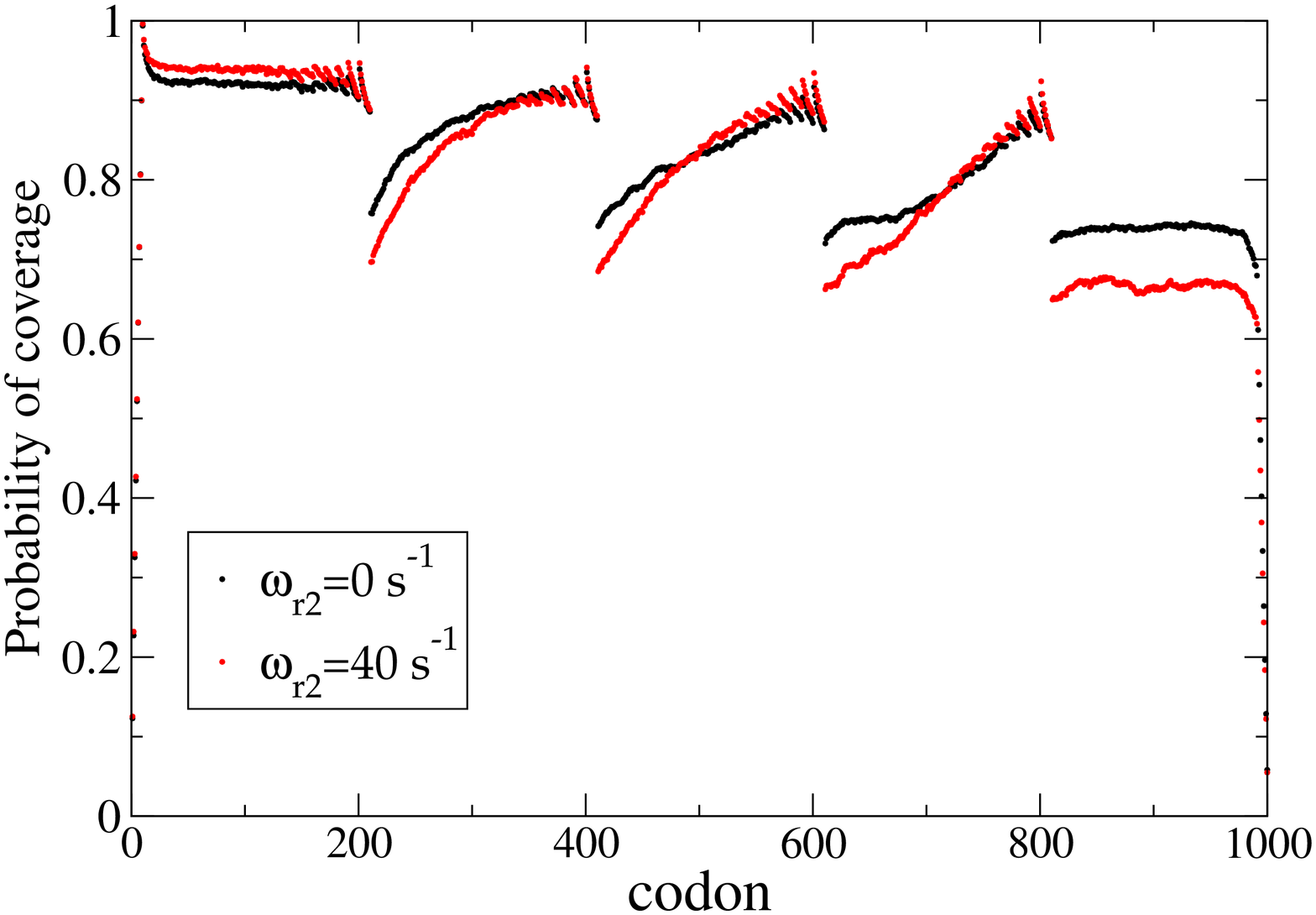}
\end{center}
\caption{} 
\label{scfigure12}
\end{figure}

\begin{figure}[ht]
\begin{center}
\includegraphics[angle=-90,width=0.75\columnwidth]{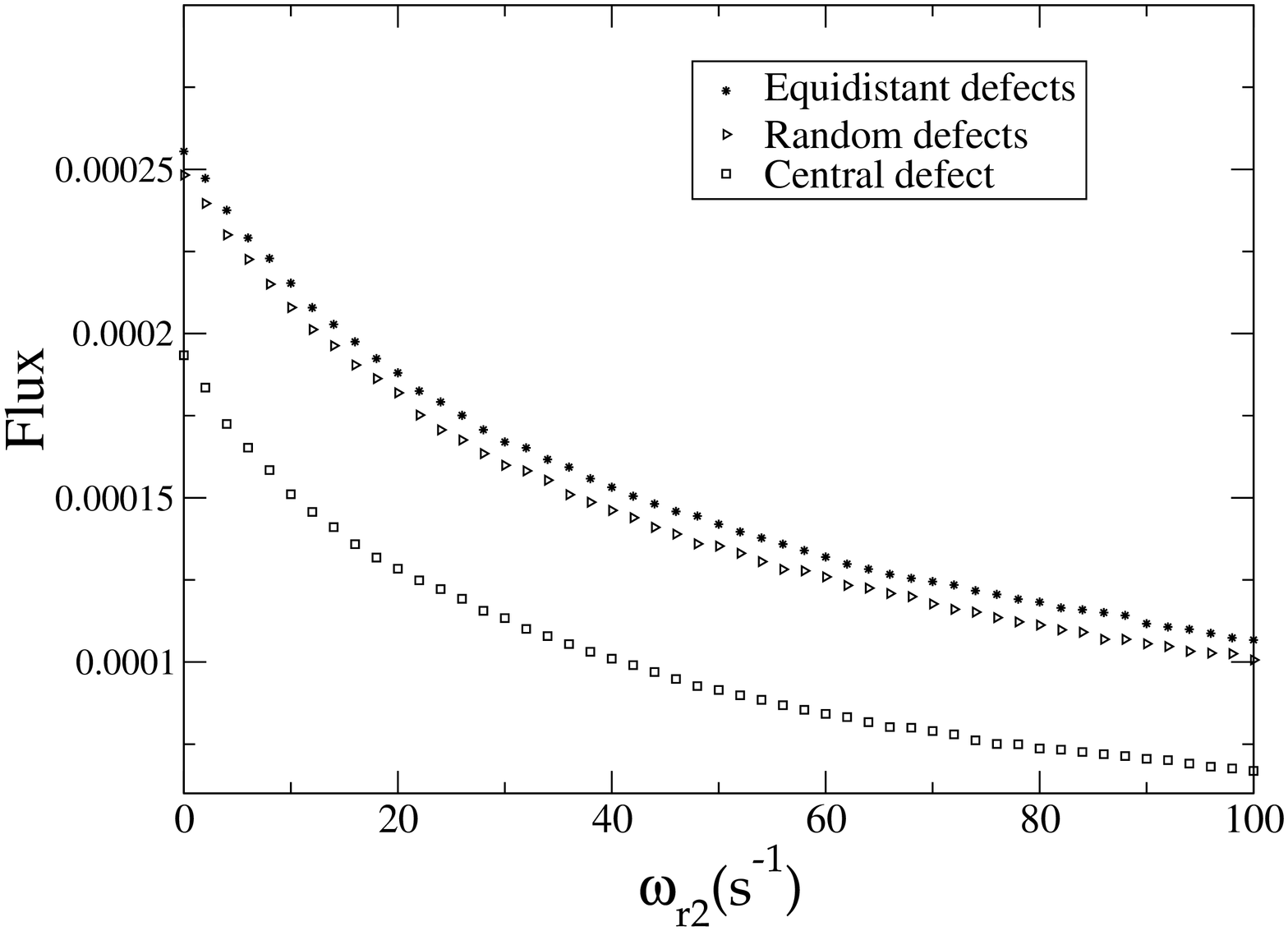}
\end{center}
\caption{} 
\label{scfigure13}
\end{figure}

\begin{figure}[ht]
\begin{center}
(a)\\
\includegraphics[width=0.75\columnwidth]{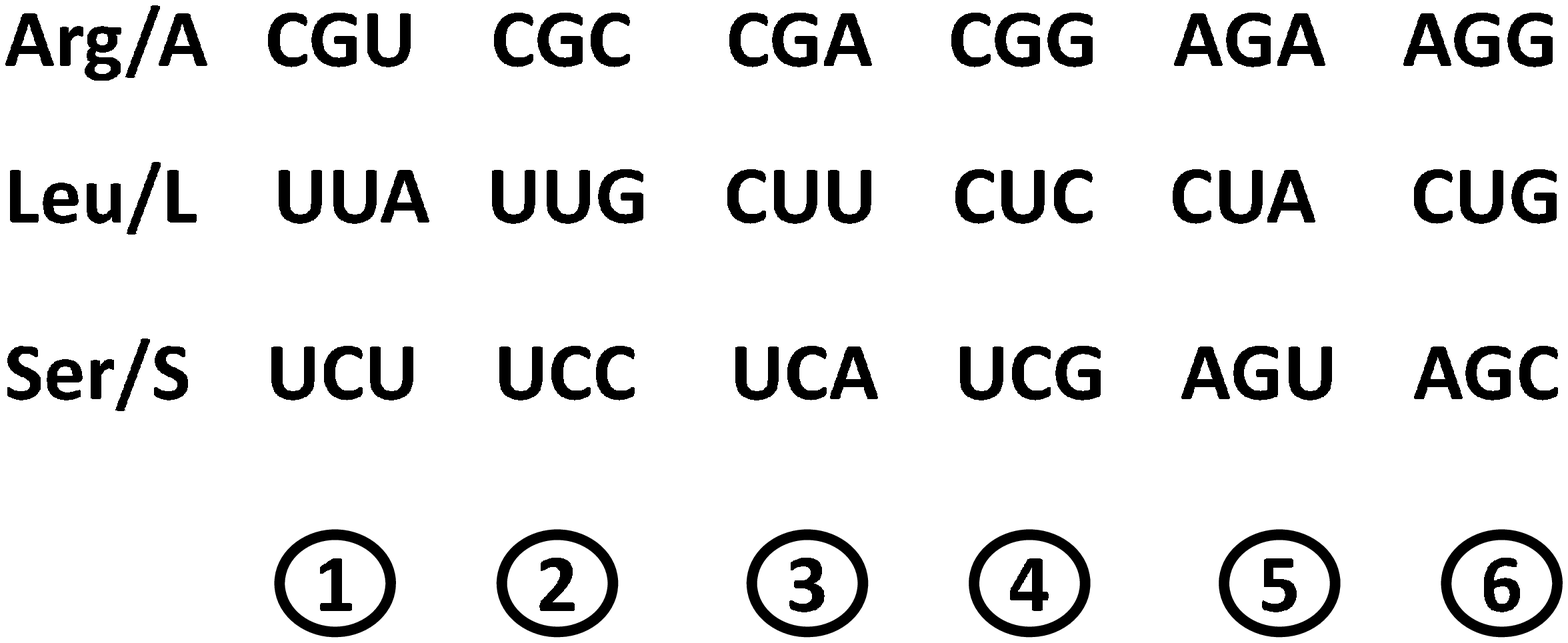}
(b) \\
\includegraphics[width=0.75\columnwidth]{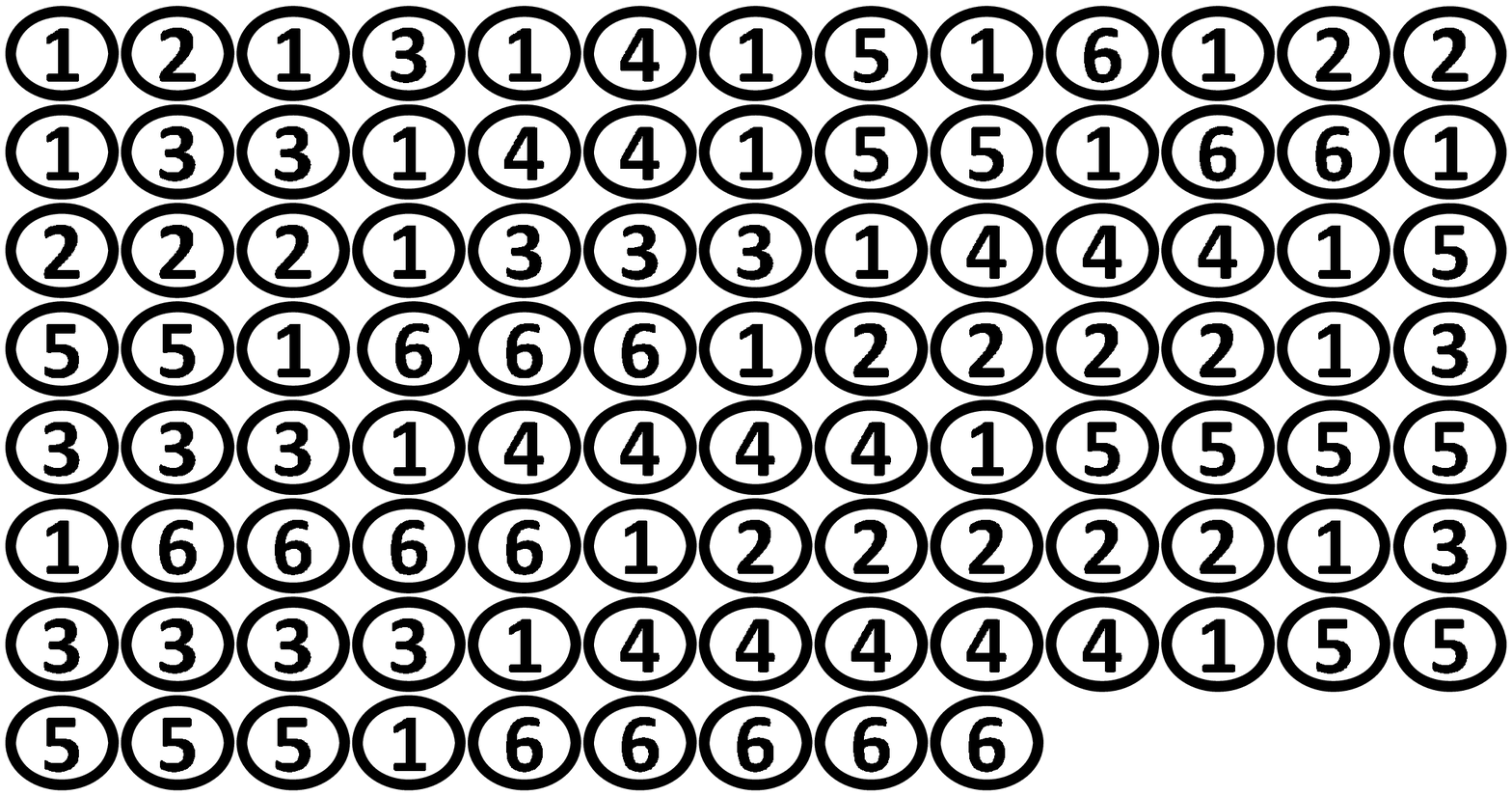}\\
\end{center}
\caption{} 
\label{fig-sequence}
\end{figure}

\noindent{\bf Fig.7}: Projections of several two-dimensional cross sections,
of the three-dimensional phase diagram, plotted in figs.\ref{fig-abr21}
and \ref{fig-abr22}, onto the $\alpha-\beta$ plane. Each
cross section corresponds to a fixed value of $\omega_{r2}$.\\

\noindent{\bf Fig.8}: 2D Phase diagram of TASEP and $\ell$-TASEP in the
$\alpha-\beta$ plane in the presence of recycling of ribosomes.\\

\noindent{\bf Fig.9}: Variation of the average flux and coverage density 
with the recycling factor $q$. The green and red curves have the same
$\alpha$ values whereas the red and blue curves have the same
$\beta$ values.\\ 

\noindent{\bf Fig.10}: Distribution of distance-headways. The lines have 
been obtained by using the formula (\ref{eq-DHfinal}) whereas the
discrete data points have been obtained directly from computer simulations.\\

\noindent{\bf Fig11}: Density profile of ribosomes for single bottleneck.\\

\noindent{\bf Fig.12}: Density profile of ribosomes for slow sites at 
$i = 200,400,600,800$.\\

\noindent{\bf Fig.13}: Flux variation with $\omega_{r2}$ for both lattices 
(see text for detail).\\ 

\noindent{\bf Fig.14}: Codon-sequence suggested for testing our theoretical 
predictions.\\

\newpage 

%%%%%%%%%%%%%%%%%%%%%%%%%%%%%%%%%%%%%%%%%%%%%%%%%%%%%%%%%%
%%%%%%%%%%%%%%%%%%%%%%%%%%%%%%%%%%%%%%%%%%%%%%%%%%%%%%%%%%

\newpage

%%%%%%%%%%%%%%%%%%%%%%%%%%%%%%%%%%%%%%%%%%%%%%%%%%%%%%%%%%
%%%%%%%%%%%%%%%%%%%%%%%%%%%%%%%%%%%%%%%%%%%%%%%%%%%%%%%%%%

\newpage 

%%%%%%%%%%%%%%%%%%%%%%%%%%%%%%%%%%%%%%%%%%%%%%%%%%%%%%%%%%
%%%%%%%%%%%%%%%%%%%%%%%%%%%%%%%%%%%%%%%%%%%%%%%%%%%%%%%%%%

\newpage 

%%%%%%%%%%%%%%%%%%%%%%%%%%%%%%%%%%%%%%%%%%%%%%%%%%%%%%%%%%
%%%%%%%%%%%%%%%%%%%%%%%%%%%%%%%%%%%%%%%%%%%%%%%%%%%%%%%%%%

\newpage 

%%%%%%%%%%%%%%%%%%%%%%%%%%%%%%%%%%%%%%%%%%%%%%%%%%%%%%%%%%
%%%%%%%%%%%%%%%%%%%%%%%%%%%%%%%%%%%%%%%%%%%%%%%%%%%%%%%%%%

\newpage 

%%%%%%%%%%%%%%%%%%%%%%%%%%%%%%%%%%%%%%%%%%%%%%%%%%%%%%%%%%
%%%%%%%%%%%%%%%%%%%%%%%%%%%%%%%%%%%%%%%%%%%%%%%%%%%%%%%%%%

\newpage 

%%%%%%%%%%%%%%%%%%%%%%%%%%%%%%%%%%%%%%%%%%%%%%%%%%%%%%%%%%
%%%%%%%%%%%%%%%%%%%%%%%%%%%%%%%%%%%%%%%%%%%%%%%%%%%%%%%%%%

\newpage 

%%%%%%%%%%%%%%%%%%%%%%%%%%%%%%%%%%%%%%%%%%%%%%%%%%%%%%%%%%
%%%%%%%%%%%%%%%%%%%%%%%%%%%%%%%%%%%%%%%%%%%%%%%%%%%%%%%%%%

\newpage 

%%%%%%%%%%%%%%%%%%%%%%%%%%%%%%%%%%%%%%%%%%%%%%%%%%%%%%%%%%
%%%%%%%%%%%%%%%%%%%%%%%%%%%%%%%%%%%%%%%%%%%%%%%%%%%%%%%%%%

\newpage 

%%%%%%%%%%%%%%%%%%%%%%%%%%%%%%%%%%%%%%%%%%%%%%%%%%%%%%%%%%
%%%%%%%%%%%%%%%%%%%%%%%%%%%%%%%%%%%%%%%%%%%%%%%%%%%%%%%%%%

\newpage 

%%%%%%%%%%%%%%%%%%%%%%%%%%%%%%%%%%%%%%%%%%%%%%%%%%%%%%%%%%
%%%%%%%%%%%%%%%%%%%%%%%%%%%%%%%%%%%%%%%%%%%%%%%%%%%%%%%%%%

\newpage 

%%%%%%%%%%%%%%%%%%%%%%%%%%%%%%%%%%%%%%%%%%%%%%%%%%%%%%%%%%
%%%%%%%%%%%%%%%%%%%%%%%%%%%%%%%%%%%%%%%%%%%%%%%%%%%%%%%%%%

\newpage 

%%%%%%%%%%%%%%%%%%%%%%%%%%%%%%%%%%%%%%%%%%%%%%%%%%%%%%%%%%
%%%%%%%%%%%%%%%%%%%%%%%%%%%%%%%%%%%%%%%%%%%%%%%%%%%%%%%%%%

\newpage 

%%%%%%%%%%%%%%%%%%%%%%%%%%%%%%%%%%%%%%%%%%%%%%%%%%%%%%%%%%
%%%%%%%%%%%%%%%%%%%%%%%%%%%%%%%%%%%%%%%%%%%%%%%%%%%%%%%%%%

\end{document}